\documentclass[traditabstract]{aa} 
\usepackage{graphicx}
\usepackage{natbib}
\usepackage{subfigure}
\usepackage{multirow}
\usepackage{txfonts}
\usepackage{longtable}

\def \kms{\ifmmode{~{\rm km\,s}^{-1}}\else{~km~s$^{-1}$}\fi}
\def \vhel{\ifmmode{V_{{\rm hel}}}\else{$V_{{\rm hel}}$}\fi}
\def \vsys{\ifmmode{V_{{\rm sys}}}\else{$V_{{\rm sys}}$}\fi}
\def \degree{\ifmmode{^{\circ}}\else{$^{\circ}$}\fi}

\def \myr{\ifmmode{{\rm\ M}_\odot{\rm\ yr}^{-1}}\else{${\rm\ M}_\odot$ 
yr$^{-1}$}\fi}

\def \mdot{\ifmmode{{\rm\dot{M}}}\else{${\rm\dot{M}}$}\fi}
\def \msun{\ifmmode{{\rm\ M}_\odot}\else{${\rm\ M}_\odot$}\fi}
\def \rsun{\ifmmode{{\rm\ R}_\odot}\else{${\rm\ R}_\odot$}\fi}
\newcommand{\HA}{H$\alpha$}
\newcommand{\HB}{H$\beta$}
\newcommand{\HBc}{H$\beta$-continuum}

\newcommand{\OIII}{[O\,{\sc iii}]\ $\lambda$5007\,\AA}

\newcommand{\SII}{[S\,{\sc ii}]\ $\lambda$6716+6731\,\AA}

\newcommand{\NII}{[N\,{\sc ii}]\ $\lambda$6584\,\AA}
\newcommand{\NIIlow}{[N\,{\sc ii}]\ $\lambda$6548\,\AA}

\begin{document}
   \title{The post-common envelope central stars of the planetary nebulae Henize~2-155 and Henize~2-161\thanks{Tables of photometry and radial velocity measurements available at the CDS via anonymous ftp to cdsarc.u-strasbg.fr}
   }
   \subtitle{}

   \author{D. Jones
          \inst{1,2}
          \and
           H.M.J. Boffin\inst{3}
                      	\and
          P. Rodr\'iguez-Gil\inst{1,2}
                     \and
                                 R. Wesson \inst{3}
                     \and
                     	R.L.M. Corradi\inst{1,2}
                     \and
           B. Miszalski\inst{4,5}
           \and
           S. Mohamed\inst{4}
                        }

   \institute{              Instituto de Astrof\'isica de Canarias, E-38200 La Laguna, Tenerife, Spain\\
  \email{djones@iac.es}
              \and
              Departamento de Astrof\'isica, Universidad de La Laguna, E-38206 La Laguna, Tenerife, Spain
                       \and
                       European Southern Observatory, Alonso de C\'ordova 3107, Casilla 19001, Santiago, Chile
                       \and
             South African Astronomical Observatory, P.O. Box 9, Observatory, 7935 Cape Town, South Africa
             \and
             Southern African Large Telescope Foundation, P.O. Box 9, Observatory, 7935 Cape Town, South Africa
             }

   \date{Received 3 December 2014 / accepted ? ??? 2015}

 
  \abstract{We present a study of Hen~2-155 and Hen~2-161, two planetary nebulae which bear striking morphological similarities to other planetary nebulae known to host close-binary central stars.  Both central stars are revealed to be photometric variables while spectroscopic observations confirm that Hen~2-155 is host to a double-eclipsing, post-common-envelope system with an orbital period of 3$^h$33$^m$ making it one of the shortest period binary central stars known.   The observations of Hen~2-161 are found to be consistent with a post-common-envelope binary of period $\sim$1 day.
  
A detailed model of central star of Hen~2-155, is produced, showing the nebular progenitor to be a hot, post-AGB remnant of approximately 0.62 M$_\odot$, consistent with the age of the nebula, and the secondary star to be an M dwarf whose radius is almost twice the expected ZAMS radius for its mass.  In spite of the small numbers, all main-sequence companions, of planetary nebulae central stars, to have had their masses and radii constrained by both photometric and spectroscopic observations have also been found to display this ``inflation''.  The cause of the ``inflation'' is uncertain but is probably related to rapid accretion, immediately before the recent common-envelope phase, to which the star has not yet thermally adjusted.

The chemical composition of both nebulae is also analysed, showing both to display elevated \emph{abundance discrepancy factors}.  This strengthens the link between elevated \emph{abundance discrepancy factors} and close binarity in the nebular progenitor. 
  }

\titlerunning{The post-CE binary central stars of Hen~2-155 \& Hen~2-161 }

   \maketitle
%

\section{Introduction} 
\label{sec:intro}

Planetary nebulae (PNe) with close binary central stars offer an important laboratory for the study of binary evolution - in particular, the poorly understood common envelope (CE) phase.  To date, approximately 50 close binary central stars (hereafter bCSPNe - binary central stars of PNe) have been discovered\footnote{http://drdjones.net/bCSPN} \citep{miszalski09a,demarco08,boffin14a} with many more expected \citep[given the high binary fraction amongst those nebulae observed by photometric surveys, e.g.\ ][]{miszalski09a,bond00}.  Finding, and accurately constraining the orbital parameters of, these missing binaries is a time consuming task, with a completely unbiased survey requiring many, many nights of photometric monitoring and subsequent spectroscopic follow-up.  As such, it is more reasonable to use the morphological features identified by \cite{miszalski09b}, as being particularly prevalent amongst PNe with binary central stars, as selection criteria for targeted photometric searches for new binaries.  These features include filamentary structures, rings, polar outflows and bipolarity, and have been successfully employed by our group in the discovery of several new binary central stars \citep[][]{miszalski11a,miszalski11b,corradi11full,boffin12b,jones14a,santander-garcia15}.

Here, we report on the discovery of post-common envelope central stars at the heart of the PNe \object{Hen~2-155} and \object{Hen~2-161}, two nebulae selected for study based on their similar appearance to other PNe known to host binary central stars. We also highlight the novel use of narrowband filters in our photometric monitoring, which afford the possibility to study some of the bright nebulae previously inaccessible to broadband photometric methods.

\object{Hen~2-155} ($\alpha = 16^h 19^m 23.17^s$, $\delta = -42^\circ 15' 36.64''$, \object{PN G338.8+05.6}) and \object{Hen~2-161} ($\alpha = 16^h 24^m 37.79^s$, $\delta = -53^\circ 22' 34.14''$, \object{PN G331.5$-$02.7})  were both discovered by Karl G.\ Henize in 1967, forming part of his catalogue of Southern planetary nebulae \citep{henize67}.  Both PNe display elongated (following the classification scheme of \citealt{sahai11}) morphologies with knotty filamentary waists, typical of those PNe already known to host binary central stars.  Beyond this initial similarity, both nebulae are remarkably similar to some of the most interesting examples in the currently known sample and were, therefore, selected by our group for study.

\begin{figure}[]
\centering
\includegraphics[width=0.35\textwidth]{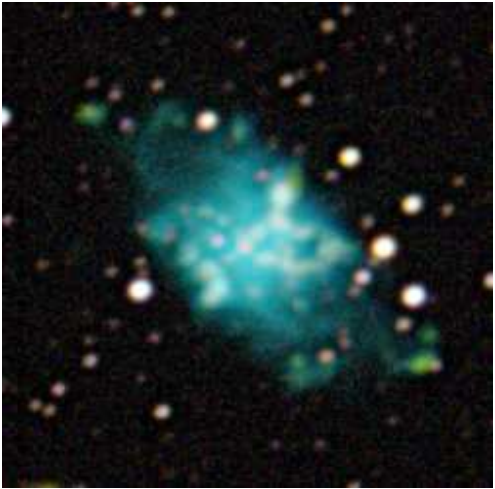}
\caption[]{Colour composite image of \object{Hen~2-155} (for exposures in individual filters please see figure \ref{fig:images}). North is up, East is to the left.  The image measures 1\arcmin$\times$1\arcmin{}.}
\label{fig:colour}
\end{figure}

The wispy extensions and irregular central filaments of \object{Hen~2-155} (see figure \ref{fig:colour}) are extremely similar to those found in \object{NGC~6326}, \object{NGC~6778}, \object{Hen~2-11}, and \object{NGC~5189} \citep{miszalski11b,jones14a,manick15}.  The geometrically opposed pairs of ``knots'' seen marking the ends of the major axis of the nebula are relatively typical of high velocity outflows or ``jets'' which are believed to be typical of central star binarity.   Of additional interest, the central binaries of NGC~6326, NGC~6778 and Hen~2-11  all display extremely strong photometric variability and periods shorter than 1 day (much shorter in the case of \object{NGC~6778} at 0.1534 day), making for relatively easy detection, whilst \object{Hen~2-11} and \object{NGC~6778} are known to host eclipsing systems.  Eclipses are key in measuring accurate and model independent masses \citep{miszalski08}, which are, in turn, extremely important in accurately constraining the state of these binaries upon exiting the CE.  The presence of the PN ensures that the system is ``fresh out of the oven'' as the nebula itself is the product of the ejection of the CE, and the lifetime of the PN is insufficient for significant changes to have occurred in the binary post-ejection. 
 
\begin{figure*}[]
\centering
\includegraphics[width=\textwidth]{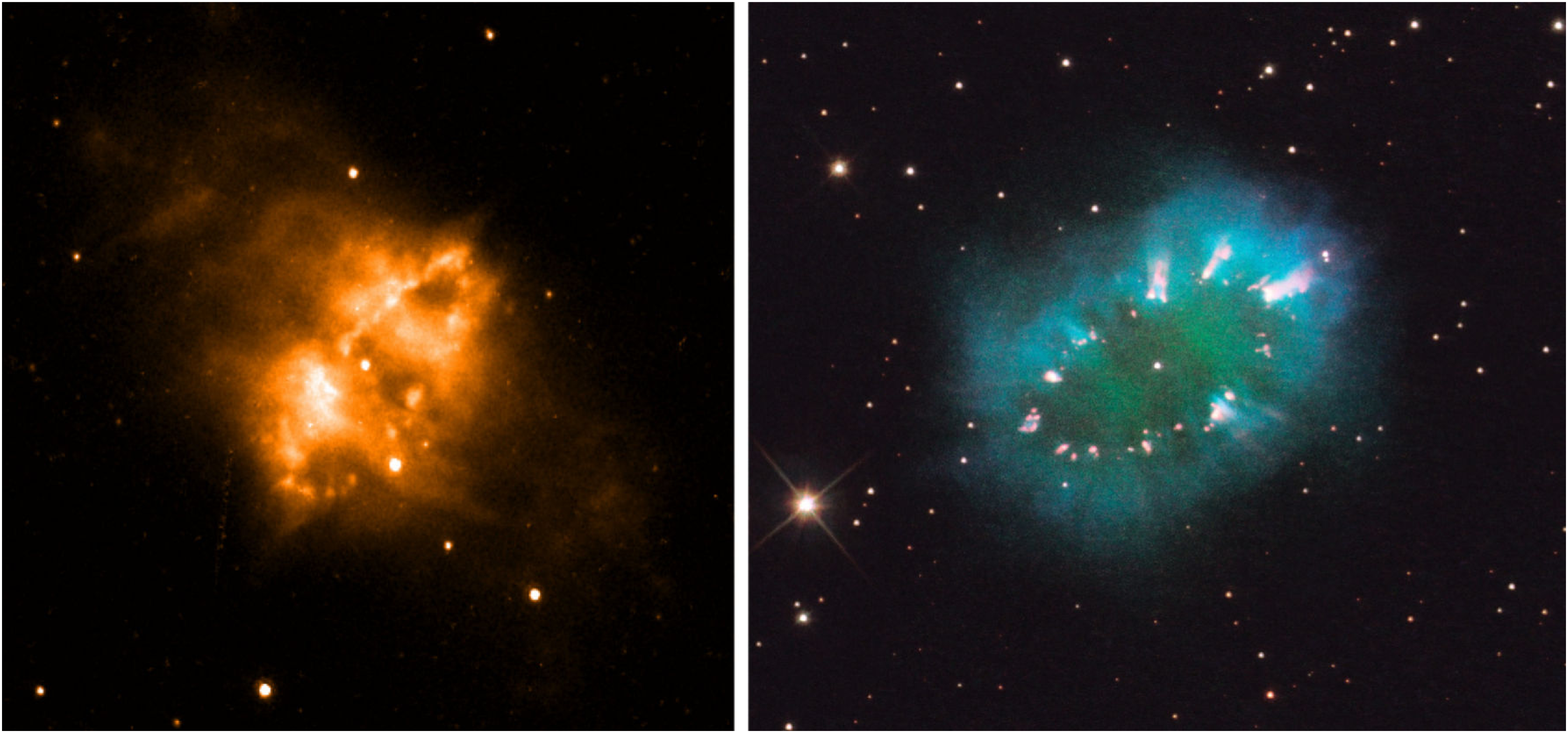}
\caption[]{HST images of \object{Hen~2-161} \citep[left, see also ][]{sahai11} and \object{The Necklace} \citep[right;][]{corradi11full} highlighting their remarkably similar appearances (elongated with knotty waists).  The image of \object{Hen~2-161} is in the light of \HA{}+\NII{}, measuring roughly 20\arcsec$\times$20\arcsec{}, while \object{The Necklace} image is a colour composite (Credit: NASA, ESA, and the Hubble Heritage Team (STScI/AURA)) measuring roughly 39\arcsec$\times$35\arcsec{}.  Note the that the central star of \object{Hen~2-161} is offset to the Northeast from the geometric centre of the nebular ring (in both images, North is up and East is to the left).}
\label{fig:161fig}
\end{figure*}

\object{Hen~2-161} shows a knotty equatorial ring forming the waist of an elongated, possibly bipolar, morphology \citep[figure \ref{fig:161fig} of this paper, and figure 3 lower panel of][]{sahai11} reminiscent of the morphology of \object{The Necklace} (figure \ref{fig:161fig}), as well as several other post-CE PNe (e.g.\ \object{Sab~41}
, \citealt{miszalski09b}; and \object{Fg~1}, \citealt{boffin12b}). 
Once again, the central binary of The Necklace shows extremely strong variability and a relatively short period \citep[1.16 day,][]{corradi11full}, but is of special interest due to the chemical peculiarity of the companion star in the system.  The secondary in \object{The Necklace} has been shown to be a Carbon Dwarf star, indicating that it has been chemically polluted (with AGB material) by the primary star - thus far the only direct evidence of mass transfer during or, much more likely, just prior to the CE phase in a PN \citep{miszalski13b}.  Additionally, the central star of \object{Hen~2-161} has been shown to be displaced from the geometric centre of the nebula \citep{sahai11}, another possible indicator of central star binarity \citep{soker98a}.

A summary of the known parameters, stellar and nebular, of both Hen~2-155 and Hen~2-161 can be found in table \ref{tab:literature}.

\begin{table}
\centering
\caption{Parameters, both stellar and nebular, of Hen~2-155 and Hen~2-161 from the literature}              
\label{tab:literature}      
\centering                                      
\begin{tabular}{r c c c}          
\hline\hline                        
 & Hen~2-155 & Hen~2-161 & References \\    
\hline
Reddening, $c$(H$_\beta$) &0.81--1.01 & 0.35--1.80 &1,2\\
Distance (kpc) & 2.47--3.15& 3.64--5.31& 3,4\\
Central star temperature&$T_Z$(H)=51& \multirow{2}{*}{$T_{EB}$=35.1}&\multirow{2}{*}{5,6}\\
 estimate(s) (kK)& $T_Z$(He~\textsc{ii})=73&&\\
Nebular systemic &\multirow{2}{*}{-35.3$\pm$15.0}&\multirow{2}{*}{-98.0$\pm$9.0}&\multirow{2}{*}{7}\\
velocity (\kms{})&&&\\
Approximate &\multirow{2}{*}{14.5--16.9}\tablefootmark{a} &\multirow{2}{*}{9.7--16.3} & \multirow{2}{*}{8}\\
nebular diameter (\arcsec{}) & &&\\
\hline
\end{tabular}
\tablebib{(1)~\citet{cavichia10}; (2)~\citet{cahn92}; (3)~\citet{phillips04}; (4)~\citet{stanghellini08}; (5)~\citet{kaler91}; (6)~\citet{preite89} ;
(7)~\citet{durand98}; (8)~\citet{tylenda03}.
}
\tablefoot{
\tablefoottext{a}{This measurement encompasses the ``waist'' of the nebula but not the wispy extensions which reach out to $\sim$30\arcsec{} from the central star.}
}
\end{table}

The paper is structured as follows.  Sect.\ \ref{sec:obs} outlines the observations and data reduction, Sect.\ \ref{sec:analysis} describes the light curve analysis and modelling, and the analysis of the stellar and nebular spectroscopy. Finally, Sect.\ \ref{sec:discussion} discusses the results.

\section{Observations and data reduction}
\label{sec:obs}
\subsection{Photometry}

\begin{figure*}[]
\centering
\includegraphics[width=\textwidth]{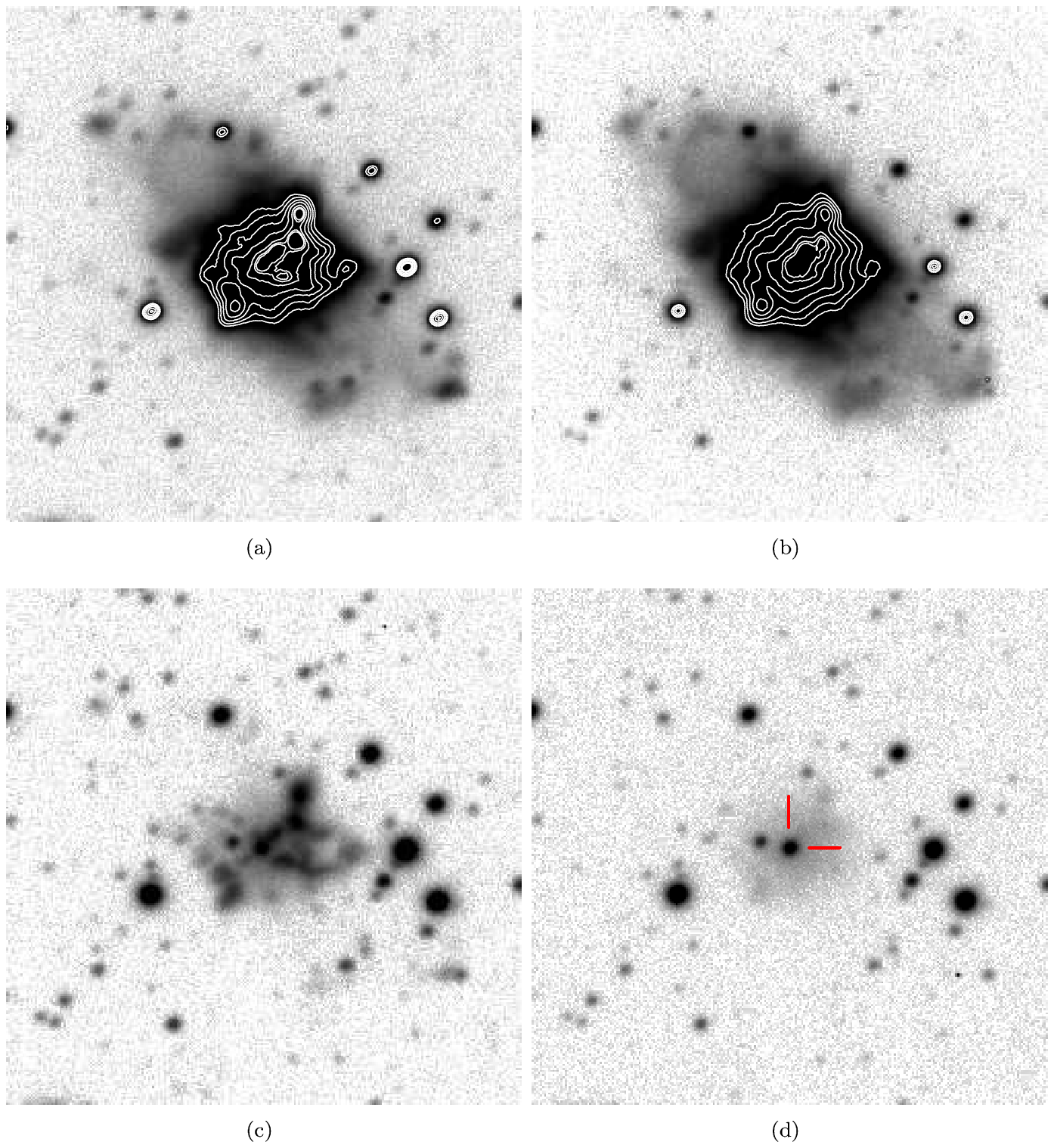}
\caption[]{NTT-EFOSC2 images of \object{Hen~2-155} in the light of (a) \HA{}+[N~\textsc{ii}], (b) \OIII{}, (c) \SII\ and (d) \HBc{} (with central star highlighted). North is up, East to the left and each image measures 1\arcmin$\times$1\arcmin{}.  All images are displayed on a logarithmic scale.  The contours in (a) and (b) represent the 99.7, 99.5, 99, 98, 97, 96 and 95$^{th}$ percentiles and highlight the brighter central structures visible in each filter.}
\label{fig:images}
\end{figure*}

Photometric monitoring of the central stars of \object{Hen~2-155} and \object{Hen~2-161} was carried out using the EFOSC2 instrument on the 3.6-m ESO-NTT \citep{EFOSC2a,EFOSC2b}.  The observations employed the E2V CCD with a pixel scale of 0.24\arcsec{}$\times$0.24\arcsec{} pixel$^{-1}$.  \object{Hen~2-155} was observed with the \HBc\ filter (\#743) on the nights of February 27--29, March 1--2 2012, and February 23--26 2014, while \object{Hen~2-161} was observed with the Gunn $i$ filter (\#705) on the nights of February 29, March 1--2 2012, January 14--15, June 2--6 2013, and February 23--26 2014.  For both objects, the exposure time was varied in order to ensure a sufficiently high signal-to-noise ratio in the resulting images (for a full list of individual exposures, see tables \ref{tab:155phot} and \ref{tab:161phot} available in the online material).  All data were debiased and flat-fielded using standard \textsc{starlink} routines\footnote{http://starlink.jach.hawaii.edu/}.

Narrowband emission line images of both objects were also acquired with EFOSC2 and the following filters: \HA\ (\# 692, includes \NIIlow{}), \OIII\ (\#687) and \SII\ (\#702), as well as in broadband Bessel B (\#639), V (\#641), R(\#642), and Gunn $i$ (\#705).  A colour composite of \object{Hen~2-155} made from the narrowband exposures is presented in figure \ref{fig:colour}, while examples of the individual exposures in each of the narrowband filters (including \HBc{}, where the central star is highlighted) are shown in figure \ref{fig:images}. 

Further images of \object{Hen~2-161} were taken using a Cousins $I$-band filter and the SAAO CCD instrument mounted on the SAAO 1.0-m telescope on the night of April 2nd 2012.  The 1\,K$\times$1\,K STE4 CCD was employed without binning, resulting in a pixel scale of 0.31\arcsec$\times$0.31\arcsec{} pixel$^{-1}$.

Contamination from high nebular background can make precision photometry of the central star difficult, even leading to spurious variability \citep[often correlated with the seeing;][]{jones11}.  As such, the choice of filter for monitoring was made based on estimates of the nebular background level with respect to the brightness of the central star in each filter.  Previous monitoring campaigns (and surveys) have experienced most success with $I$-band monitoring, as this offers the best combination of broad bandpass (to reduce exposure times) and avoids the brightest nebular emission lines.  However, for brighter nebulae with similarly bright central stars (like \object{Hen~2-155}), even the $I$-band can show marked nebular emission, in these cases, monitoring can be successfully performed in \HBc{}.  The \HBc{} filter (\#743) excludes \emph{all} nebular emission lines reducing the nebular contamination purely to continuum emission, offering the opportunity to observe the central star with minimal influence from the nebula.  Given the large aperture of the ESO-NTT and the bright nature of the central star of \object{Hen~2-155}, photometric monitoring was possible in \HBc{} with high signal-to-noise ratio and with minimal nebular contamination.

Even with the use of filters excluding the majority of the nebular emission, some remains, as such photometry was extracted from both targets with an aperture tailored to a diameter of roughly 3$\times$ the maximum seeing during the observations \citep[5\arcsec{}, thus minimising the variable contamination in the aperture;][]{jones11}.  Photometry was performed using the \textsc{sextractor} software \citep{bertin96}, and the differential magnitude of the central stars measured against non-variable field stars.  The $I$-band observations of \object{Hen~2-161} were then placed on an absolute scale using catalogue photometry from DENIS \citep{DENIS} and the methodology described in \cite{boffin12a}, with an approximate precision of 0.05 mag (derived from the dispersion of detector zero points calculated from each field star).  The \HBc{} observations of \object{Hen~2-155} were also placed on an approximate absolute scale using standard stars observed during the observing runs (this can only be considered approximate due to the variable nature of extinction during each night and each run, however as the value of each data point is determined by relative brightness to field stars no spurious variability can have been introduced as part of this process).  All photometric measurements and their uncertainties are shown in tables \ref{tab:155phot} and \ref{tab:161phot} available in the online version, and in machine-readable format via anonymous FTP from the CDS.  

\addtocounter{table}{2}

\subsection{Spectroscopy}

\subsubsection{Hen~2-155}
\label{sec:hen2155spec}

The central star of \object{Hen~2-155} was observed on the night of March 19 2013 with FORS2 mounted on the ESO-VLT's Antu telescope \citep{FORS}.  The instrument set-up consisted of a longslit measuring 0.5\arcsec{}$\times$6.8\arcmin{}, the GRIS\_1400V grism and a mosaic of two unbinned 4k$\times$2k MIT/LL CCDs, offering a spectral resolution of roughly 0.7\AA{} with a spectral range of $\sim$4560--5860\AA{}.  The observations consisted of 13 contiguous exposures of 900-s followed by a further exposure of 1200-s.  The frames
were de-biased and flat-fielded in {\sc iraf}\footnote{{\sc iraf} is
distributed by the National Optical Astronomy Observatories.}. We then
used \textsc{pamela} \citep{marsh89} to remove the sky contribution
and obtain the 1D spectra by means of optimal extraction
\citep{horne86} as implemented in \textsc{pamela}.

The wavelength scale was obtained from one arc spectrum taken during the
FORS2 daytime standard calibration routines. We checked for any
instrument flexure by monitoring the stability of the nebular [O\,{\sc
iii}] $\lambda$4959 emission line. This nebular line was slowly shifting
with time, so we removed this trend from each individual spectrum
accordingly. This wavelength calibration process and the radial velocity
analysis presented below were carried out in \textsc{molly}\footnote{Tom
Marsh's {\sc molly} package is available at 
http://deneb.astro.warwick.ac.uk/phsaap/software/molly/html/INDEX.
html}.

The nebula of \object{Hen~2-155} was observed again with the VLT-FORS2 on the night of March 14 2014, with an exposure time of 1200-s employing the GRIS\_1200B grism (3660 \AA{}$<\lambda<$5110\AA{}), directly followed by a 120-s exposure taken with the GRIS\_1200R grism and a GG435 filter (5750 \AA{}$<\lambda<$7310\AA{}).  Both spectra were taken using a 0.7\arcsec{}$\times$6.8\arcmin{} slit (at a position angle of 40 \degr{}, along the major axis of the nebula) and binning of 2$\times$2 (providing a resolution of approximately 1.5\AA{}).  Standard \textsc{starlink} routines were employed to bias subtract, flat field correct, wavelength and flux calibrate the spectra \citep[flux calibration was performed using observations of the standard star LTT4816 taken on the same nights and using the same instrumental setups; ][]{hamuy92}.

\subsubsection{Hen~2-161}

The central star of \object{Hen~2-161} was observed on 31 May 2012 with the queue-scheduled Southern African
Large Telescope \citep[SALT; ][]{buckley06b,odonoghue06}. 
The Robert Stobie Spectrograph \citep[RSS; ][]{burgh03,kobulnicky03} was used with the PG1300 grating and 1.5\arcsec{}$\times$8\arcmin{} slit
to obtain a 2150-s exposure covering $\lambda=4278-6360$ \AA, at a mean resolution of 4.1\AA\ and a reciprocal
dispersion of 0.66\AA\ pixel$^{-1}$.  
Basic reductions were applied using the PySALT\footnote{http://pysalt.salt.ac.za} package \citep{crawford10}. Cosmic ray events were cleaned using the \textsc{lacosmic} package
\citep{vandokkum01}. Wavelength calibration of the contemporaneous Argon arc lamp exposure was
performed using standard \textsc{IRAF} tasks \textsc{identify},
\textsc{reidentify}, \textsc{fitcoords} and \textsc{transform} by
identifying the arc lines in each row and applying a geometric
transformation to the data frames. The one dimensional spectrum of the central star
was extracted using \textsc{apall} and no nebular subtraction was made.

VLT-FORS2 observations of \object{Hen~2-161} and its central star were acquired on the night of March 14 2014, with an exposure time of 120-s, employing the GRIS\_1200R grism and a GG435 filter , and on April 1 2014, with an exposure time of 1140-s and the GRIS\_1200B grism (for further details of the setup see section \ref{sec:hen2155spec}). All spectra were taken using a 0.7\arcsec{}$\times$6.8\arcmin{} slit (at a position angle of 45\degr{}, chosen to cover the major axis of the nebula).  Standard \textsc{starlink} routines were employed reduce the spectra just as described in section \ref{sec:hen2155spec}.

\section{Analysis}
\label{sec:analysis}

\subsection{Hen~2-155}
\subsubsection{Lightcurve and radial velocities}

\begin{figure*}[]
\centering
\includegraphics[width=\textwidth]{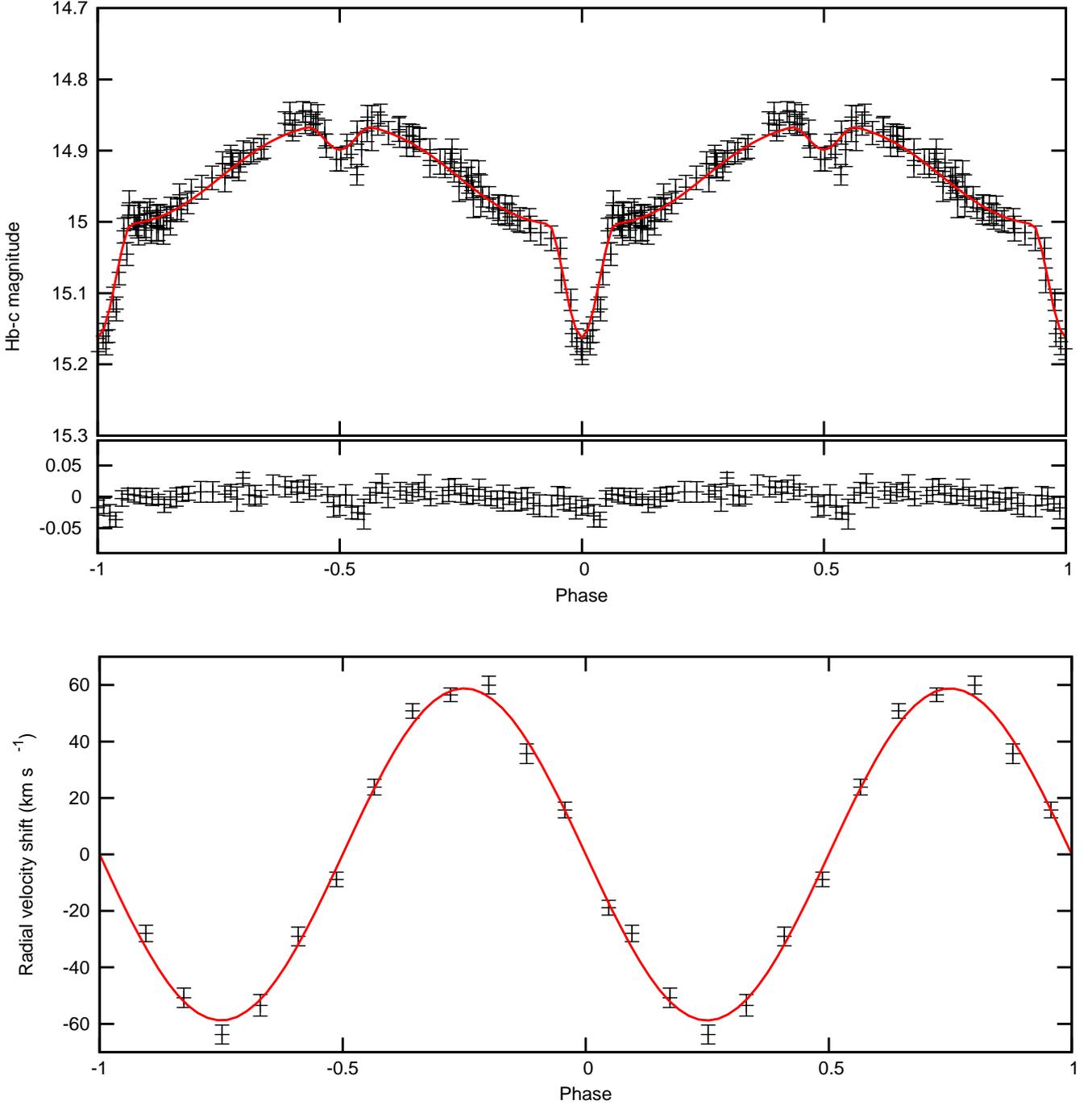}
\caption[]{Folded NTT-EFOSC2 \HBc{} photometry of the central star of \object{Hen~2-155} with the \textsc{nightfall} model overlaid in red (upper panel).  Binned residuals between the observed photometry and model are shown in the middle panel.  Note that the size of the points represents the photometric uncertainty and contain no estimate of the uncertainty due to variable nebular contamination (which be significant, at certain phases).  The lower panel shows the radial velocity measurements of the central star of \object{Hen~2-155} with the predicted radial velocity curve from the \textsc{nightfall} model overlaid in red.}
\label{fig:phot155}
\end{figure*}

A Lomb-Scargle analysis was performed in order to determine the periodicity of the variability displayed by the central star of \object{Hen~2-155} using the \textsc{period} package of the \textsc{starlink} software suite \citep{period}.  Fig.\ \ref{fig:phot155} shows the data folded on the ephemeris determined by the analysis,

$$\mathrm{min}\;\mathrm{HJD}= 2\,455\,985.7865(\pm 0.0001) + 0.148275 (\pm 0.000008) E$$

where min~HJD is Heliocentric Julian Date of the \HBc{} lightcurve minimum.

The lightcurve shows smooth sinusoidal variation of peak-to-peak amplitude $\sim$0.15 mag, with eclipses at phase 0 and 0.5.  The primary eclipse (as the secondary passes in front of the primary), at phase 0, is approximately 0.2 mag deep, while the secondary eclipse measures approximately 0.05 mag deep.  
The ``v-shaped'' eclipse profiles are typical of grazing eclipses, indicating a moderate orbital inclination (high inclinations would lead to total eclipses which have a ``flat-bottomed'' profile).  Given that at least one of the stars should be a post-AGB remnant (having ejected the PN), then the overall sinusoidal variability is likely to arise from the varying projection of the face of the secondary which is being heated by the hot post-AGB star (known as an ``irradiation effect'' or ``reflection effect'').  Given the short period, this variability could also be attributed to one or both stars being close to filling their Roche lobes and, as such, presenting a variable surface area throughout the orbit (known as ``ellipsoidal modulation''), this effect is usually associated with rarer, double-degenerate binaries \citep[where both components are post-AGB;][]{hillwig10,bruch01,santander-garcia15}.  However, ``ellipsoidal modulation'' presents two peaks per period with any eclipses occurring at the minima, therefore as the curve shows only one peak per period the origin of the overall variability must originate from the irradiation of the secondary by the primary.

The spectra of Hen~2-155 (Fig.\ \ref{fig:spec155}) show absorption lines associated with a hot, post-AGB star (e.g.\ O~\textsc{v} $\lambda$5114\,\AA{}) as well as high-excitation emission lines from the nebula (e.g.\  [Ar \textsc{iv}] $\lambda$4711+4740\,\AA{}).  The spectra also show emission lines which have previously been identified as typical of an irradiated main-sequence secondary star (C~\textsc{iv} $\lambda$5801+5812\,\AA{}).

Radial velocity measurements, of the primary component of the central star of \object{Hen~2-155}, were made via a cross-correlation of the individual O~\textsc{v} $\lambda$5114\,\AA{} absorption profiles in the VLT-FORS2 spectra, and using the average profile as a template.
In producing this template, the
radial velocities of the O~\textsc{v} $\lambda$5114\,\AA{} line were first measured by fitting
Gaussians to the individual profiles. This preliminary radial velocity
curve was then fit with a sine function and the measured orbital motion removed from each individual spectrum using this sine fit.
Finally, the resulting profiles were averaged  in order to
obtain the template for cross-correlation. This procedure provides a
cross-correlation template with good signal-to-noise ratio and a line
width representative of that of the individual profiles. The final measured radial velocities (shown folded in the lower panel of Fig.\ \ref{fig:phot155} and listed in the online table \ref{tab:155rv}) agree extremely well with the ephemeris determined from the photometry, clearly showing that the photometric period is also the orbital period.  The variation is well fit by a sinusoid of amplitude 59.6$\pm$1.5 \kms{}, indicating an extremely low orbital eccentricity (as would be expected for such a short period system).  The relatively low amplitude (given the short period) of the variability strongly indicates that either: the orbital inclination is low (as the amplitude varies with the sine of the inclination), or, the mass ratio, $q$, is small (as the amplitude varies with $(1+q)^{^{-2}/_3}$).  As the lightcurve does show shallow eclipses, it is most likely a combination of the two effects that results in the observed amplitude.

\begin{figure}[]
\centering
\includegraphics[width=\columnwidth]{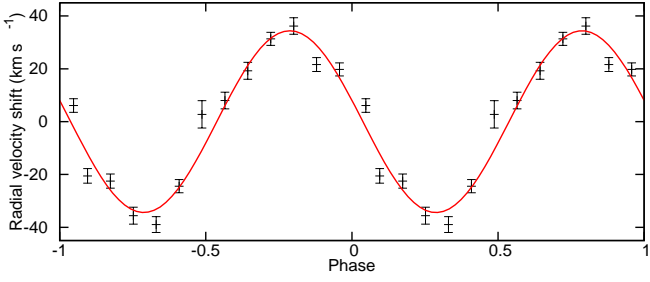}
\caption[]{Radial velocity curve from the C~\textsc{iv} lines in \object{Hen~2-155}.}
\label{fig:CivRV}
\end{figure}

\begin{figure}[]
\centering
\includegraphics[width=\columnwidth, trim = 0 0 0 0, clip]{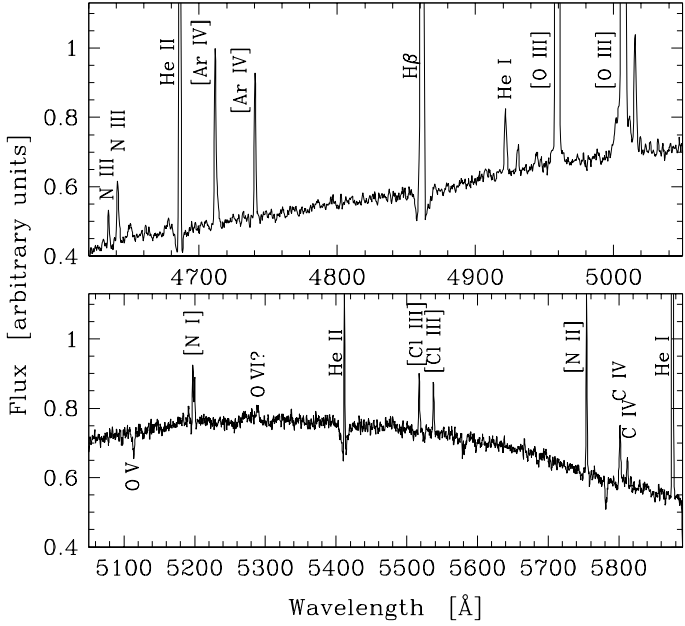}
\caption[]{An example VLT-FORS2 spectrum of \object{Hen~2-155} and its central star. 
}
\label{fig:spec155}
\end{figure}

All other lines visible in the spectra, including emission lines which may be associated with the secondary of the system (e.g.\ N~\textsc{iii} $\lambda$4634+4641\,\AA{} and C~\textsc{iv} $\lambda$5801+5812\,\AA{}), were also inspected for radial velocity shifts.  All  but the C~\textsc{iv} complex at $\sim$5800\,\AA{} were found to be non-variable or too weak to be measured (e.g. the possible detection of O~\textsc{vi} $\lambda$5291\,\AA{}).  These lines were found to move roughly in phase with the O~\textsc{v} absorption (a phase shift of 0.037 is calculated from the fit, shown in figure \ref{fig:CivRV}) but with a lower amplitude ($\sim$35 \kms{} c.f. $\sim$60 \kms{}).  The radial velocities measured from the C~\textsc{iv} lines, using the same cross-correlation technique as for the O~\textsc{v} $\lambda$5114\,\AA{} absorption feature (described above), are shown folded on the orbital period, along with a sinusoidal fit, in figure \ref{fig:CivRV} and listed in the online table \ref{tab:155rvCiv}.  The lower amplitude appears difficult to reconcile with the measured radial velocities of the O~\textsc{v} absorption feature, which must originate from the surface of the primary star.  The most likely explanation is that these lines form in the wind of the hot primary and as such trace a region above the WD surface but which has to be restricted to a zone more-or-less between the two stars (as emission from an isotropic wind would present the same radial velocity variability as lines originating from the stellar surface).  That the emission is restricted to a relatively small region may be indicative of a ``hot spot'' on the surface of the primary \citep{deschamps13}.

\begin{figure}[]
\centering
\includegraphics[width=\columnwidth, trim = 20 40 340 320, clip]{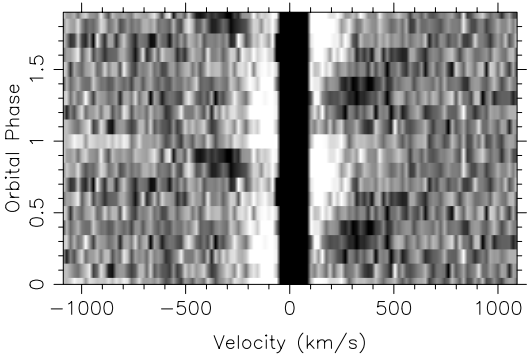}
\caption[]{Trailed spectra showing the emission around the He~\textsc{ii} $\lambda$4686\,\AA{} nebular line. Emission is shown in black, and absorption in white.}
\label{fig:trailed155}
\end{figure}

An unusual profile is found around the He~\textsc{ii} $\lambda$4686\,\AA{} nebular emission line (the trailed spectra are shown around this line in figure \ref{fig:trailed155}.  As is common in spectra of CSPN, the bright nebular emission line is still visible (even after a basic background subtraction, due to its bright and irregular nature) on top of a broad absorption profile moving in phase with the O~\textsc{v} absorption line and with a compatible amplitude (we do not measure its amplitude due to the obvious complication of the nebular emission), but at phases of approximately 0.25 and 0.75 there is evidence of high velocity emission in anti-phase with the lines originating from the primary.  Given the strong contamination from both the nebula and primary, it is not possible to disentangle this emission, and therefore determine its origin and possible association with the secondary.

\addtocounter{table}{2}

To properly constrain the parameters of the system, simultaneous modelling, of both the radial velocities (from the O~\textsc{v} line) and lightcurve, was performed using the \textsc{nightfall} code\footnote{http://www.hs.uni-hamburg.de/DE/Ins/Per/Wichmann/Nightfall.html}.  All parameters were varied over a wide range of physical solutions, with the final model being selected for having the lowest $\chi^2$ fit.  Detailed reflection was employed in the modelling (with 5 iterations) in order to properly treat the irradiation of the secondary by the primary.  A model atmosphere\footnote{http://kurucz.harvard.edu/grids/gridP00/ip00k2.pck19} 
was used for the lower temperature (secondary) component with solar metallicity and log $g$ of 4.5 \citep{kurucz93}.  The final model lightcurve is shown, along with the residuals to the binned data, in Fig.\ \ref{fig:phot155} and the model parameters outlined in Tab.\ \ref{params}.

\begin{table}
\centering
\caption{Modelled and observed binary parameters for \object{Hen~2-155}}              
\label{params}      
\centering                                      
\begin{tabular}{r c c}          
\hline\hline                        
 & Primary & Secondary \\    
\hline                                   
$T_\mathrm{eff}$ (K) & 90\,000$\pm$5\,000 & 3\,500$\pm$500 \\
Radius ($R_\odot$) & 0.31$\pm$0.02 & 0.30$\pm$0.03\\
Log $g$& -- & 4.5\tablefootmark{a} \\
\hline
Inclination & \multicolumn{2}{c}{68.8\degr{}$\pm$0.8\degr{}}\\
$q=^{M_2}/_{M_1}$ & \multicolumn{2}{c}{0.21$^{+0.05}_{-0.02}$}\\
$M_{tot}$ ($M_\odot$) &  \multicolumn{2}{c}{0.75$\pm$0.05}\\
Period (day) & \multicolumn{2}{c}{$0.148275\pm$0.000008}\\
\hline                                             
\end{tabular}
\tablefoot{
\tablefoottext{a}{A fixed parameter in the modelling}
}
\end{table}

\begin{figure}[]
\centering
\includegraphics[height=\columnwidth,angle=270]{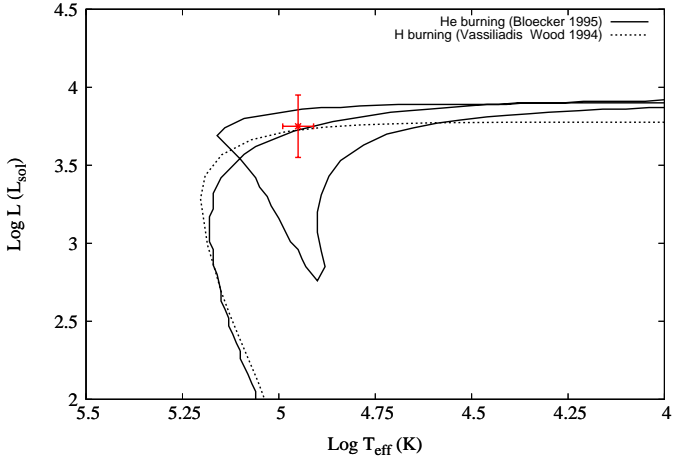}
\caption[]{The temperature and luminosity of the model central star of Hen~2-155 overlaid on top of evolutionary tracks for a remnant of mass 0.625 M$_\odot$ taken from \citet{bloecker95} and 0.623 M$_\odot$ taken from \citet{vassiliadis94}.}
\label{fig:evolplot}
\end{figure}

The model represents an excellent fit, with residuals smaller than the uncertainties of individual observations for almost all data points (the scatter of the radial velocity measurements around the model curve is greater than the photometric data, but still generally a good fit). The modelled progenitor star lies on the stellar evolutionary tracks of both \citet[][with an He burning remnant]{bloecker95} and \citet[][an H burning remnant with initial mass, M$_i$=1.0 M$_\odot$, and metallicity, $Z$=0.001]{vassiliadis94} for a remnant mass of 0.625 M$_\odot$ and 0.623 M$_\odot$, respectively, both in excellent agreement with the modelled mass of 0.62 M$_\odot$ (see figure \ref{fig:evolplot}).  The post-AGB age is roughly 10\,000 years for both of these tracks, which is a reasonable age for a planetary nebula and, for a distance of $\sim$3 kpc \citep{zhang95,phillips04,stanghellini08}, would imply an expansion velocity of $\sim$30 \kms{} \citep[again, perfectly reasonable for a PN, e.g.;][]{jones10a,jones12}.   It is worth noting that the model parameters also lie close to tracks from both authors with different parameters (mass, H or He burning, metallicity, etc.), giving an age range from $\sim$3,000 years up to $\sim$25,000 years.  Neither end of this expansive range would be out of the question for a PN, however they would imply an exceptionally large or particularly low expansion velocity at the lower and upper age limits, respectively.  This range is perhaps reflective of the uncertainty in the starting point for post-AGB tracks, as well as the fact that this is a post-CE system (rather than a single star evolving off the AGB with no external influences, as considered by the models).

The model luminosity (defined by the Boltzmann law, and the model radius and temperature) gives an apparent $V_{0}$ for the central star of $\sim$14.7 mag  \cite[assuming a reasonable bolometric correction of $-7$;][]{reed98}.  From the single $i$-band observation taken at a phase of 0.67 (we choose not to use the $V$-band observation due to the extremely high levels of nebula contamination), we measure an extinction-corrected $V_{0}$ of $\sim$14.4 mag (uncorrected $V_{0}$ of 15.6 mag)\footnote{Extinction correction performed assuming the extinction law of \citet{howarth83}, a $c(H\beta)$ of 0.74 (see section \ref{sec:ionic155}), and $(V-I)_O=-0.40$ \citep{ciardullo99}.}.  At first this would appear inconsistent, however, noting that the contribution from the secondary is near maximum at this phase, the non-negligible nebular contamination, as well as the uncertainties on the assumed values (particularly the distance), the two values are most certainly consistent and provide a valuable check of the model.  

The model temperature of the primary star is, at 90 kK, in relatively poor agreement with those in the literature determined by the Zanstra method \citep[51$\pm$10 kK and 73$\pm$10 kK for  He~\textsc{i} and He~\textsc{ii} Zanstra temperatures, respectively;][]{kaler91}.  This is to be expected because the Zanstra method depends on accurate measurement of broadband central star magnitudes which will, of course, be contaminated by the contribution from the companion star \citep[as seen in][]{jones14a}. Furthermore, the Zanstra temperature determinations assume an optically thick nebula, which would not necessarily be the case for such a hot central star.  As such, the Zanstra temperatures can only be considered lower limits for the true central star temperature.  The poor agreement between the Zanstra temperatures calculated from Hydrogen and Helium emission lines, further indicates that these temperatures may not be reflective of the true central star temperature.  \citet{kaler91} found a strong correlation between the intensity of the nebular \OIII\ line and the central star temperature, and via their analytical relationship the Stoy Temperature of the central star would be expected to be $\sim$90,000 K, in good agreement with the modelled temperature.  Furthermore, the presence of O~\textsc{vi} emission, in the VLT-FORS2 spectrum of the central star, indicates a high stellar temperature, probably around $\sim$100,000 K (much higher than this and the O~\textsc{v} lines become much less pronounced, lower and the O~\textsc{vi} line would not be present).

The modelled secondary mass and temperature of 0.13 M$_\odot$ and 3\,500 K are in rough agreement with those predicted by ZAMS models for an M5V star \citep[the ZAMS temperature of a 0.13 M$_\odot$ star with solar metallicity is approximately 3\,000 K;][]{MESA}, here the slightly elevated temperature could be partly due to the uncertainty on the model temperature as well as the fact that the high levels of irradiation from the primary may cause a global increase in temperature rather than just an increase on the ``day side'' of the secondary.  The modelled radius, however, is much greater than expected for a main sequence star of this mass (almost a factor of two larger, more in keeping with a star of mass 0.3 M$_\odot$).  This ``inflation'' could be associated with the high levels of irradiation from the primary, but could also be a symptom of mass transfer just before entering the CE phase.  Further discussion of this ``inflation'' shall be reserved for Section \ref{sec:discussion}.

The orbital plane, at an inclination of 68.8\degr{}$\pm$0.8\degr{}, is expected to be perpendicular to the nebular symmetry axis \citep{jones14b,nordhaus06}, while the nebula appears to be viewed more or less side-on (implying an inclination of $\sim$90\degr{} for the binary orbital plane).  However, it is not possible to determine the nebular orientation without detailed spatio-kinematic modelling (which necessitates the acquisition of high-resolution, spatially-resolved, nebular spectroscopy) and, moreover, PNe with similar projections have been found to have inclinations that would be generally consistent with the modelled inclination \citep[for example, HaTr~4 appears similarly side-on but, in fact, has an inclination of $\sim$75\degr{};][]{tyndall12}.

\subsubsection{Ionic and total abundances of Hen~2-155}
\label{sec:ionic155}

\begin{figure}[]
\centering
\includegraphics[height=\columnwidth,angle=270]{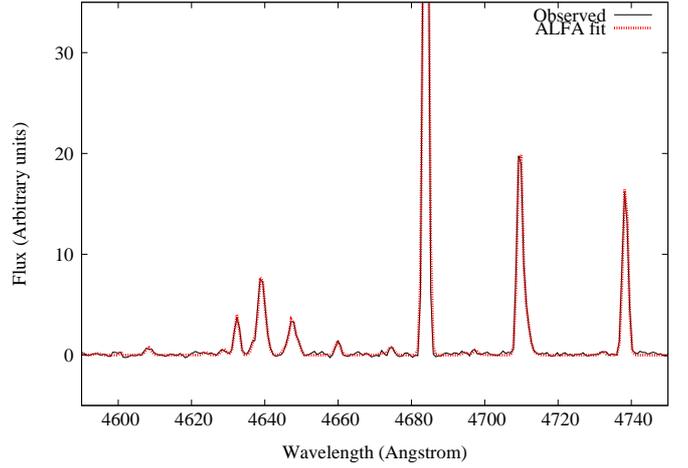}
\caption[]{The continuum-subtracted observations and \textsc{alfa} fit for a subsection of the FORS2 GRIS\_1200B spectrum of the nebula Hen~2-155, covering the C~\textsc{iii} and N~\textsc{iii} recombination lines as well as several lines of the V1 O~\textsc{ii} multiplet, which are critical in determining the nebular \emph{adf}.}
\label{fig:alfa}
\end{figure}

Apertures containing nebular emission from Hen~2-155 were extracted from the VLT-FORS2 spectroscopy described in section \ref{sec:obs}.  From the resulting one-dimensional spectra, emission line fluxes were derived using the \textsc{alfa} code (Wesson, in prep.; Jones et al.\ in prep.), which optimises the parameters of Gaussian fits to the line profiles using a genetic algorithm, fitting all lines simultaneously after subtracting a globally-fitted continuum (see figure \ref{fig:alfa} for a demonstration of the quality of the fit provided by \textsc{alfa}).  These emission line fluxes were then used to derive abundances using the \textsc{neat} code \citep{wesson12}.  This code corrects for interstellar extinction using measured Balmer line ratios, and then determines temperatures and densities from the standard diagnostics.  Elemental abundances are then calculated as the flux-weighted averages of the emission lines from each species, and total chemical abundances using the ionisation correction scheme of \citet{delgado-inglada14}.  At each stage, uncertainties are robustly propagated using a monte carlo technique.

Measured and dereddened fluxes (along with their 1$\sigma$ uncertainties) for Hen~2-155  are shown in Table \ref{tab:fluxes}, while the determined nebular properties are listed in table \ref{tab:etd}.  The reddening is found to be $c$(H$_\beta$)=0.74$\pm$0.01 in relatively good agreement with the values found in the literature (0.81, \citealt{cavichia10}; 0.98--1.01, \citealt{cahn92}), as are the individual line fluxes c.f.\ those presented by \citet{cavichia10}.

The final ionic and total abundances for Hen~2-155 are shown in tables \ref{tab:ionic} and \ref{tab:abund}, respectively.  The abundances indicate that Hen~2-155 is a non-type \textsc{i} PNe \citep[based on the He/H and N/O ratios;][]{peimbert78,kingsburgh94}.  Furthermore, comparison of Oxygen abundances determined from collisionally excited lines (CELs) and optical recombination lines (ORLs), show a strong discrepancy -- known as an abundance discrepancy factor (\emph{adf}) -- whereby the abundance from ORLs is a factor of 6 greater than that from CELs.  Such discrepancies are well known, appearing in analyses of all photoionised nebulae, but are generally found to be only a factor of a few \citep[1.5--3 in H~\textsc{ii} regions, while only $\sim$20\% of PNe are found to have \emph{adfs} $>$ 5;][]{garcia-rojas07,wesson05}.  Further discussion of this enhanced \emph{adf} is reserved for section \ref{sec:discussion}.

\addtocounter{table}{1}

\begin{table}
\centering
\caption{Extinction, temperatures, and densities as determined by the empirical analysis using the \textsc{neat} code.}              
\label{tab:etd}      
\centering                                      
\begin{tabular}{r c c}          
\hline\hline                        
 & Hen~2-155 & Hen~2-161 \\    
\hline
$c$(H$_\beta$) & 0.74$\pm$ 0.01& 1.21$\pm$0.01\\
$T_e$([N~\textsc{ii}]) (K) & -- & 9800$\pm$360\\
$T_e$([O~\textsc{iii}]) (K) & 11660$\pm$40 &8190$\pm$110\\
$n_e$([O~\textsc{ii}]) (cm$^{-3}$)&1300$\pm$70 &1870$\pm$110\\
$n_e$([S~\textsc{ii}]) (cm$^{-3}$)& 1390$\pm$55&1500$\pm$89\\
$n_e$([Ar~\textsc{iv}]) (cm$^{-3}$)&670$\pm$160&--\\
\hline
\end{tabular}
\end{table}

\begin{table}
\centering
\caption{Ionic abundances, relative to Hydrogen (in the form $\frac{ion}{H}$), as determined by the empirical analysis using the \textsc{neat} code.  All abudances are calculated using CELs unless otherwise indicated.}              
\label{tab:ionic}      
\centering                                      
\begin{tabular}{r c c}          
\hline\hline                        
 Ion & Hen~2-155 & Hen~2-161 \\    
\hline
C$^{2+}$ (ORLs)                          & ${  3.39\times 10^{ -4}}^{+  1.9\times 10^{ -5}}_{ -1.8\times 10^{ -5}}$  & ${  2.13\times 10^{ -3}}^{+  4.4\times 10^{ -5}}_{ -4.3\times 10^{ -5}}$\\
C$^{3+}$ (ORLs)                        & ${  1.51\times 10^{ -4}}^{+  1.7\times 10^{ -5}}_{ -1.6\times 10^{ -5}}$ &  ${  2.75\times 10^{ -3}}^{+  3.4\times 10^{ -4}}_{ -3.4\times 10^{ -4}}$ \\
N$^{+}$                           & ${  5.27\times 10^{ -6}}^{+  1.8\times 10^{ -8}}_{ -1.7\times 10^{ -8}}$ & ${  7.89\times 10^{ -6}}^{+  8.1\times 10^{ -7}}_{ -7.3\times 10^{ -7}}$\\
O$^{+}$                           & ${  1.22\times 10^{ -5}}^{+  1.2\times 10^{ -7}}_{ -1.2\times 10^{ -7}}$ & ${  1.52\times 10^{ -5}}^{+  2.7\times 10^{ -6}}_{ -2.3\times 10^{ -6}}$\\
O$^{2+}$ (CELs)                       & ${  1.43\times 10^{ -4}}^{+  2.9\times 10^{ -6}}_{ -2.8\times 10^{ -6}}$ & ${  2.05\times 10^{ -4}}^{+  1.2\times 10^{ -5}}_{ -1.1\times 10^{ -5}}$ \\
O$^{2+}$ (ORLs)                     & ${  9.01\times 10^{ -4}}^{+  3.8\times 10^{ -5}}_{ -3.6\times 10^{ -5}}$ & ${  2.26\times 10^{ -3}}^{+  8.5\times 10^{ -5}}_{ -8.2\times 10^{ -5}}$ \\
adf(O$^{2+}$) & 6.3 & 11.0\\
Ne$^{2+}$                       & ${  5.07\times 10^{ -5}}^{+  6.7\times 10^{ -7}}_{ -6.6\times 10^{ -7}}$& ${  7.75\times 10^{ -5}}^{+  5.2\times 10^{ -6}}_{ -4.9\times 10^{ -6}}$ \\
Ar$^{2+}$                         & ${  8.76\times 10^{ -7}}^{+  4.7\times 10^{ -8}}_{ -4.5\times 10^{ -8}}$& ${  1.66\times 10^{ -6}}^{+  6.5\times 10^{ -8}}_{ -6.2\times 10^{ -8}}$ \\
Ar$^{3+}$                       & ${  3.91\times 10^{ -7}}^{+  5.6\times 10^{ -9}}_{ -5.5\times 10^{ -9}}$ & ${  1.87\times 10^{ -7}}^{+  2.4\times 10^{ -8}}_{ -2.2\times 10^{ -8}}$\\
S$^{+}$                           & ${  2.44\times 10^{ -7}}^{+  2.0\times 10^{ -9}}_{ -2.0\times 10^{ -9}}$ & ${  2.76\times 10^{ -7}}^{+  2.6\times 10^{ -8}}_{ -2.4\times 10^{ -8}}$\\
S$^{2+}$                          & ${  1.83\times 10^{ -6}}^{+  5.7\times 10^{ -8}}_{ -5.6\times 10^{ -8}}$& ${  2.50\times 10^{ -6}}^{+  3.4\times 10^{ -7}}_{ -3.0\times 10^{ -7}}$ \\

\hline
\end{tabular}
\end{table}

\begin{table}
\centering
\caption{Total nebular abundances from Hen~2-155 and Hen~2-161, determined using the \textsc{neat} code.}              
\label{tab:abund}      
\centering                                      
\begin{tabular}{llll}
\hline\hline
Element & Hen~2-155 & Hen~2-161\\
\hline
H & 1.00 & 1.00 \\
He& ${  0.11}^{+6.5\times 10^{ -4}}_{-6.5\times 10^{ -4}}$ &${  0.13}^{+  7.7\times 10^{ -4}}_{ -8.3\times 10^{ -4}}$\\
C (ORLs) & ${  4.91\times 10^{ -4}}^{+  2.5\times 10^{ -5}}_{ -2.4\times 10^{ -5}}$ & ${  4.87\times 10^{ -3}}^{+  3.5\times 10^{ -4}}_{ -3.3\times 10^{ -4}}$ \\
N & ${  7.01\times 10^{ -5}}^{+  1.5\times 10^{ -6}}_{ -1.4\times 10^{ -6}}$&${  1.14\times 10^{ -4}}^{+  8.9\times 10^{ -6}}_{ -8.2\times 10^{ -6}}$  \\
O & ${  1.62\times 10^{ -4}}^{+  3.0\times 10^{ -6}}_{ -2.9\times 10^{ -6}}$&${  2.21\times 10^{ -4}}^{+  1.2\times 10^{ -5}}_{ -1.1\times 10^{ -5}}$  \\
Ne &${  5.74\times 10^{ -5}}^{+  6.7\times 10^{ -7}}_{ -6.6\times 10^{ -7}}$&${  8.35\times 10^{ -5}}^{+  5.4\times 10^{ -6}}_{ -5.1\times 10^{ -6}}$  \\
S & ${  3.50\times 10^{ -6}}^{+  1.1\times 10^{ -7}}_{ -1\times 10^{ -7}}$& ${  4.81\times 10^{ -6}}^{+  6.7\times 10^{ -7}}_{ -5.9\times 10^{ -7}}$ \\
Ar &${  1.26\times 10^{ -6}}^{+  4.8\times 10^{ -8}}_{ -4.6\times 10^{ -8}}$&${  1.84\times 10^{ -6}}^{+  7.8\times 10^{ -8}}_{ -7.5\times 10^{ -8}}$ \\
\hline
\end{tabular}
\end{table}
 
 \subsubsection{Photoionisation modelling of Hen~2-155}

In order to confirm that a hot (T$\sim$90 kK) central star, predicted by our modelling of the photometric light curve, is consistent with the overall ionisation state of the nebula, we constructed a simple photoionisation model for comparison to our measured emission line fluxes.  It is important to note that our model is only intended to show that the ionisation state of the nebula is consistent with the central star temperature determined from the simultaneous modelling of light- and radial velocity curves.  It is not intended to provide a realistic description of the nebula, given that the nebula is clearly both irregular and filamentary (Fig.\ \ref{fig:images}) and that we have no information about surface abundances in the central star with which to realistically estimate the ionising spectrum.

\begin{figure}[]
\centering
\includegraphics[angle=270,width=\columnwidth]{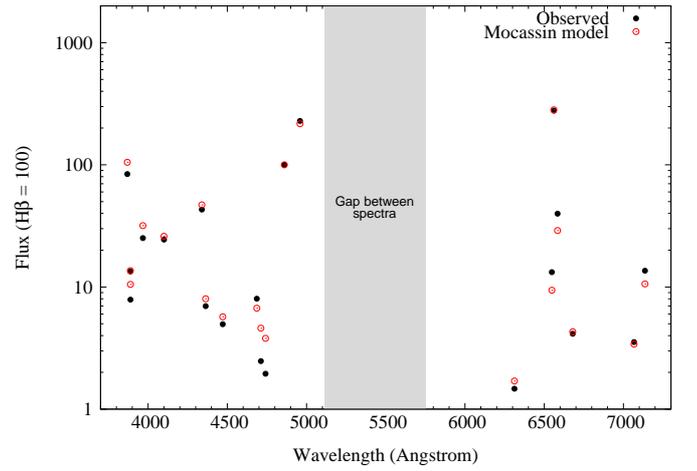}
\caption[]{A comparison of emission line fluxes from the \textsc{mocassin} model of Hen~2-155 and those measured from the FORS2 spectroscopy (Table \ref{tab:fluxes}).}
\label{fig:mocassinmodel}
\end{figure}

We modelled the nebula, using \textsc{moccasin} \citep{ercolano03}, as a sphere of radius $10^{18}$ cm, estimated from the angular extent of the nebula and a distance of 3~kpc.  The model has uniform hydrogen density of 1000 cm$^{-3}$ (see previous section), and abundances based on the empirical analysis of observed line fluxes (for a list of the final abundances used see table \ref{tab:abund}).  The model reproduces fairly well the overall emission spectrum of the nebula (see figure \ref{fig:mocassinmodel}) despite being so simplistic.  The central star in the model is a blackbody with a temperature of 90~kK and a luminosity of 5200~L$_\odot$, both matching well with the temperature and luminosity determined from simultaneous modelling of the light- and radial velocity curves.  A more realistic spectral shape for the ionising source (i.e.\ not a perfect blackbody) and more realistic density distribution would certainly affect the ionisation balance, but the model clearly shows that a hot central star is consistent with the measured emission line ratios.  As such, we conclude that the temperature determined in this work, 90~kK, constitutes a more accurate estimate of central star temperature than those made using other methodologies in the literature (e.g.\ Zanstra temperatures).

\subsection{Hen~2-161}
\subsubsection{Stellar photometry and spectroscopy}

The photometric observations of \object{Hen~2-161} were taken in two different filters - the majority in Gunn-$i$ taken using NTT-EFOSC2 with the rest taken in Cousins-$I$ with the SAAO CCD mounted on the SAAO 1.0-m. In order to ensure no colour effects were present in the combination of data taken through two different $I$-band filters, the relative magnitude of several field stars (used in photometering the central star) were measured, with the results generally agreeing within the uncertainties of the individual measurements.  As such, we do not apply any colour correction to the photometry of the central star - this may, indeed, induce spurious variability, so the results will be assessed in the context of the combined data and the NTT data alone (as the vast majority of the data originate from this telescope).

\begin{figure}[]
\centering
\includegraphics[width=\columnwidth]{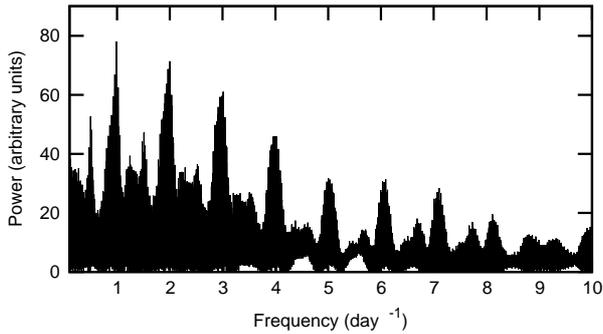}
\caption[]{Lomb-Scargle periodogram of the photometric observations of \object{Hen~2-161}.}
\label{fig:lombscargle161}
\end{figure}

\begin{figure*}[]
\centering
\includegraphics[width=0.8\textwidth]{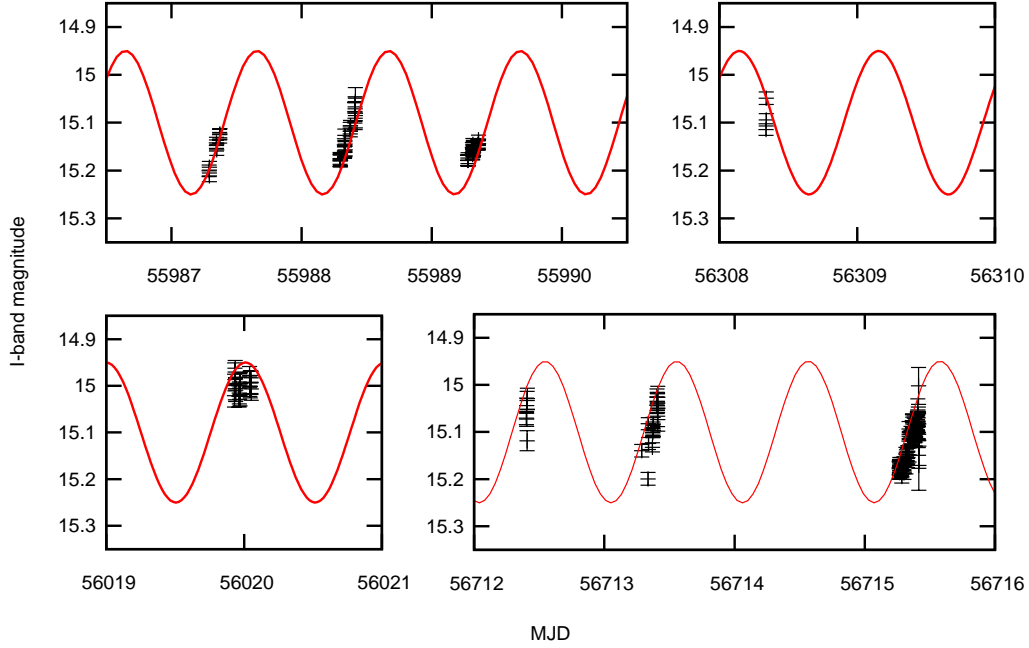}
\caption[]{$I$-band photometry of the central star of \object{Hen~2-161} with an overlaid sinusoid of period 1.011 day (the peak of the periodogram shown in figure \ref{fig:lombscargle161}).  The observations from the SAAO 1.0-m telescope (taken in a different $I$-band filter to the rest of the observations) are shown in the lower left panel, highlighting that even without adding the data from this telescope, the variability is still clear.  The scatter in the points may be due to an error in the period or to variable nebula contamination as seen in \citet{miszalski11b}.}
\label{fig:phot161}
\end{figure*}

A Lomb-Scargle analysis was performed for the complete photometry of the central star of \object{Hen~2-161} using the \textsc{period} package of the \textsc{starlink} software suite \citep{period}.  The data show clear variability, but unfortunately any true period remains difficult to ascertain most likely due to it being extremely close to the periodic aliases of nightly observing (12 hours and 1 day).  The Lomb-Scargle periodogram of the data is presented in figure \ref{fig:lombscargle161}, highlighting the strength of each of the peaks around each alias.  The data show a maximum variation of roughly 0.2 mag (0.15 mag excluding the SAAO 1.0-m data), although, given the possible periods, this is not certain to be the true peak-to-peak amplitude of the variability.  It is highly unlikely that this variability is spurious, given that it does not correlate with any aspect of the observing parameters other than the time of the observation \citep[spurious variability due to nebular contamination or colour effects generally can be found to correlate with the seeing or airmass at the time of the observation;][]{jones11}.  Even if the inclusion of the SAAO 1.0-m data is erroneous, there is still a clear upward trend in the data obtained with EFOSC2-NTT (see figure \ref{fig:phot161}).

\begin{figure}[]
\centering
\includegraphics[width=\columnwidth]{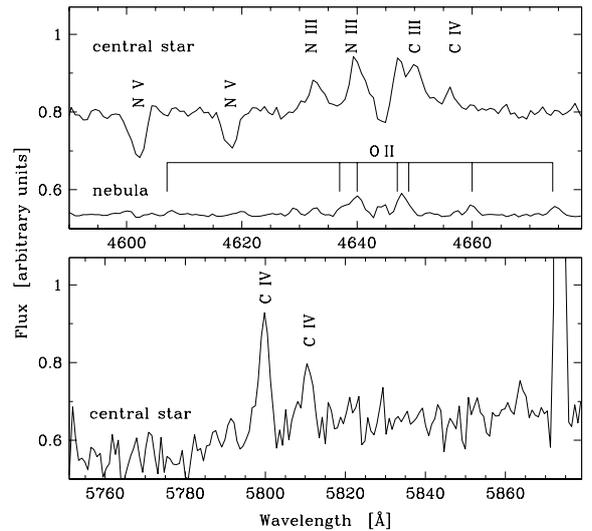}
\caption[]{VLT-FORS2 spectroscopy of the central star of \object{Hen~2-161} displaying both absorption features from a hot post-AGB central star and emission line complexes frequently associated with an irradiated, main-sequence secondary.  In both panels, the central star spectrum is shown without any nebula subtraction.  In the upper panel, an extraction of the nebula close to the central star is displayed (with an arbitrary offset) in order to highlight that the emission lines originate from the central star and not from the nebula.}
\label{fig:spec161}
\end{figure}

The VLT-FORS2 spectra and SALT-RSS spectrum all show absorption features associated with a hot post-AGB nebular progenitor (e.g.\ He~\textsc{ii} $\lambda$4542\,\AA{}, He~\textsc{ii} $\lambda$5412\,\AA{}, N~\textsc{v} $\lambda$4604\,\AA{}) as well as emission lines frequently, but not exclusively (see above discussion of these lines in the spectra of Hen~2-155), associated with irradiated, main-sequence companions \citep[see figure \ref{fig:spec161}, and][]{miszalski11b,corradi11full}.  Both absorption and emission features visible in all spectra were inspected for velocity shifts, with respect to nebular emission lines (e.g.\ \HB{}, He~\textsc{i} $\lambda$4472\,\AA{}, [O\,\textsc{iii}] $\lambda$4959\,\AA{}), with all radial velocities agreeing within the uncertainties of the measurements (which, given the low resolution of the spectrum, are rather large for this purpose).  This is, perhaps, not unreasonable given that, for a period of $\sim$1 day, the spectra would have all taken at around the same phase which may have been close to conjunction where any radial velocity shifts would be minimal (this is consistent with the  phases of the highly speculative sine curve overlaid in fig \ref{fig:phot161} where all spectra would have been taken at phase $\sim$0).  However, this does mean that we cannot rule out that the variability of the central star is not associated with binarity, but, given the persistent nature of the observed photometric variability \citep[pulsations often vary on timescales shorter than the 2 years over which our observations span, particularly in such a young post-AGB object;][]{winget08}, the presence of emission lines associated with an irradiated secondary, and the off-centre central star \citep[often associated with central star binary;][]{soker01,jones10b}, we conclude that \object{Hen~2-161} is most likely host to a close binary central star with a possible orbital period of approximately 1 day.  We strongly encourage follow-up observations, particularly at other longitudes, in order to confirm and further constrain its nature as a post-CE binary.

\subsubsection{Ionic and total abundances of Hen~2-161}

Just as for Hen~2-155 (Section \ref{sec:ionic155}), nebular apertures of Hen~2-161 were extracted from the VLT-FORS2 spectroscopy described in section \ref{sec:obs} and processed using the \textsc{alfa} and \textsc{neat} codes.  The measured and dereddened fluxes (along with their 1$\sigma$ uncertainties) for Hen~2-161  are shown in Table \ref{tab:fluxes}, while the determined properties of the nebula are listed in table \ref{tab:etd}. The reddening is found to be $c$(H$_\beta$)=1.21$\pm$0.01, while the literature values range from 0.35--1.80 \citep{cahn92}, depending on the method used in its determination (the upper value is determined from the ratio of radio 5 GHz flux density and H$\beta$, while the lower is from the ratio of H$\alpha$ and H$\beta$).  Given that our measurement is the weighted average of all the Balmer line ratios available in the covered wavelength range (which covers H$\alpha$ through to H$\zeta$), and the results are consistent across all ratios (as reflected by the relatively small uncertainty), it is clear that the value measured here must represent the most accurate determination of the reddening to date.

The final ionic and total abundances are shown in tables \ref{tab:ionic} and \ref{tab:abund}, respectively.  Hen~2-161 shows systematically higher heavy element abundances than Hen~2-155, just about satisfying the general criterion for Type \textsc{i} PNe \citep[He/H$\geq$0.125, N/O$\geq$0.5][]{peimbert78}\footnote{Note that \citet{kingsburgh94} proposed a higher threshold for being Type \textsc{i} which Hen~2-161 does not satisfy (N/O$\geq$0.8)}.  This may be indicative of a higher mass progenitor or that the binary evolution has had little (or lessened) effect on the chemical evolution of the primary \citep{demarco09}.  However, just as for Hen~2-155, Hen~2-161 displays a large \emph{adf} (over 10), possibly typical of central star binarity (see section \ref{sec:discussion}).

\section{Discussion}
\label{sec:discussion}

\begin{table*}
\centering
\caption{Binary CSPNe from the literature where parameters for both primary and secondary stars have been derived from simultaneous modelling of photometric and radial velocity observations.  The upper five have main-sequence companions, while the lower two are shown to be double-degenerates.}              
\label{tab:bins}      
\centering                                      
\begin{tabular}{r c c c c c c c l} 
Nebula & Period & M$_{CS}$ & R$_{CS}$ & T$_{CS}$ & M$_S$ & R$_S$ & T$_S$ & Notes\\
& (day) & (M$_\odot$) & (R$_\odot$) & (kK) & (M$_\odot$) & (R$_\odot$) & (kK) &\\
\hline \hline
\object{Abell~46}\tablefootmark{a} & 0.47 & 0.51$\pm$0.05 & 0.15$\pm$0.02 & 49.5$\pm$4.5 & 0.15$\pm$0.02 & 0.46$\pm$0.02 & 3.9$\pm$0.4 &Double-lined\\
\object{Abell~63}\tablefootmark{a} & 0.47 & 0.63$\pm$0.05 & 0.35$\pm$0.01 & 78$\pm$3 & 0.29$\pm$0.03 & 0.56$\pm$0.02 & 6.1$\pm$0.2 &Double-lined\\
\object{Abell~65}\tablefootmark{b} & 1.00 & 0.56$\pm$0.04 & 0.056$\pm$0.008 & 110$\pm$10 &0.22$\pm$0.04 &0.41$\pm$0.05 & 5.0$\pm$1.0 & Not eclipsing\\
\object{BE Uma}\tablefootmark{c} & 2.29 & 0.70$\pm$0.07 & 0.08$\pm$0.01 & 105$\pm$5 & 0.36$\pm$0.07 & 0.72$\pm$0.05 & 5.8$\pm$0.3 & Double-lined\\
\object{Ds 1}\tablefootmark{d} & 0.36 & 0.63$\pm$0.03 & 0.16$\pm$0.01 & 77$\pm$3 & 0.23$\pm$0.01 & 0.40$\pm$0.01 & 3.4$\pm$1 & Not eclipsing, double-lined\\
\hline
\object{Hen~2-428}\tablefootmark{e} & 0.18 & 0.88$\pm$0.13 & 0.68$\pm$0.04 & 32.4$\pm$5.2 &0.88$\pm$0.13 & 0.68$\pm$0.04 & 30.9$\pm$5.2 & Double-lined\\
\object{NGC~6026}\tablefootmark{f} & 0.53 & 0.57$\pm$0.05 & 1.06 $\pm$0.05& 38$\pm$3 & 0.57$\pm$0.05 & 0.05$\pm$0.01 & 146$\pm$15 & -\\
\hline
\end{tabular}
\tablebib{
(a) \citet{afsar08}; (b) \citet{hillwig15};
(c) \citet{ferguson99}; (d) \citet{hilditch96};
(e) \citet{santander-garcia15}; (f) \citet{hillwig10}.
}
\end{table*}

The central star of \object{Hen~2-155} is a photometrically variable binary star with a period of 0.148 day, while the central star of \object{Hen~2-161} has also been shown to be photometrically variable with a possible period of $\sim$1 day.  While a $\sim$1 day period is highly suspicious, one other bCSPN is known to have a period extremely close to $\sim$1 day \citep[Abell~65, ][]{shimansky09,bond90} meaning that such a period should not be ruled out.  However, it is ultimately unclear whether the photometric variability in the central star of \object{Hen~2-161} is associated with binarity, but we conclude that it is the most likely cause and encourage follow-up observations (both photometric and spectroscopic) at in order to fully reconstruct any orbit.  Given its brightness ($I_{}\sim$15.1 mag) and southern declination, it should be easily accessible with available telescopes, the main constraint comes in the form of seeing as, due to a bright star approximately 2\arcsec{} away from the CSPN (to the South, more or less on the edge of the nebular ring, see figure \ref{fig:161fig}), good seeing is needed to ensure photometry of the central star alone.

The central star of \object{Hen~2-155} was also found to be a single-lined, radial-velocity variable with a period and ephemeris consistent with that derived from the photometric observations.  Simultaneous modelling of the lightcurve and radial velocity curve shows that the system consists of a hot nebular progenitor, with a temperature of $\sim$90 kK, and a much cooler companion.  The modelled primary mass, radius and temperature are consistent with evolutionary tracks and the ionisation state of the nebula.  The modelled secondary has a much larger radius, and slightly higher temperature, than expected for its mass.  It is worthy of note, that there is an indication of emission from the secondary star in the system, but that its radial velocity in this line could not be measured due to low signal-to-noise and contamination from spectral features of the primary and nebula.

Radial velocity measurements of the C~\textsc{iv} $\lambda$5801+5812\,\AA{} emission lines present in the spectra of Hen~2-155 show variability in phase with that of the O~\textsc{v} absorption lines originating from the primary, but with a lower amplitude.  The true origin of this emission is quite uncertain, but it is most likely to originate from the wind of the hot primary and hence trace a region above its surface. It's restriction to a region between the two stars may be associated with a ``hot spot'' on the surface of the primary from which the wind is enhanced \citep{deschamps13}.  A ``hot spot'' would be in keeping with the Roche-lobe filling nature of the secondary, whereby direct impact of mass transferred from the secondary onto the surface of the primary would be responsible for its formation (no accretion disk would be formed because the primary is larger than the circularisation radius for any material transferred onto it).  Given the faintness of these lines, and that no other lines display the same behaviour, we conclude that while of great interest and warranting further investigation, this unusual line profile gives no reason to doubt the stellar parameters derived from the modelling.

Only a handful of other binary CSPNe have had their parameters derived by in-depth studies similar to the one presented here \footnote{Note that other systems have also been subjected to simultaneous modelling of spectroscopic and photometric observations but, due to their non-eclipsing nature (and the lack of multi-band photometry), their parameters (especially radii) could not be derived \citep[e.g.\ HFG~1;][]{exter05}. Similarly, other systems have been modelled based on photometry alone, but here masses can only be estimated via indirect means \citep[e.g.\ Hen~2-11;][]{jones14a}.  It is also worth noting that the majority of the other systems listed in table \ref{tab:bins} are double-lined spectroscopic binaries, meaning that a more direct constraint can be placed on their mass ratios than in the case of a single-lined binary (as is the case for Hen~2-155, where the parameter is constrained by the modelling).}  (with the results summarised in table \ref{tab:bins}).  Of those seven systems, two are found to be double-degenerates where both stars have evolved to the post-AGB phase, while the other five are found to have secondaries which are greatly inflated with respect to the ZAMS radii for their given masses.

The origin of the inflation seen in the secondaries of CSPNe is unclear, but several possibilities have been considered in the literature.  \citet{afsar08} considered that the inflated secondaries in \object{Abell~46} and \object{Abell~63} may originate from magnetic activity, ruling this possibility out due to the also increased temperatures (magnetic activity has been shown to result in increased radii and reduced temperatures).  \citet{demarco08} state that the extreme heating of the secondaries by the much hotter primaries could lead to them ``puffing up'' as the star is irradiated by leakage of radiation from the ``day side'' to the ``night side'', the level of this effect is, however, unclear and may not contribute sufficiently to increase the radii by the levels seen.  This hypothesis is, however, supported by the elevated temperatures determined both from the modelling and from spectroscopic observations of the ``night side'' of the secondaries \citep{demarco08}.  Finally, the inflation could be associated with mass from the primary being accreted just prior to or during the CE phase, knocking the secondary out of thermal equilibrium.  For fully convective secondaries (or those with deep convective envelopes), the amount of matter required to be accreted in order to produce the observed inflation is minimal, but the accretion rate does need to be significant \citep[$\gtrsim 10^{-5}$ M$_\odot$\,yr$^{-1}$;][]{prialnik85}.  Hydrodynamic models indicate that the spiral-in speeds during the CE are highly supersonic, leading to the formation of a bow shock and preventing any material from being accreted by the secondary during the CE phase \citep{sandquist98}.  However, recently it has been suggested that there could be a period of pre-CE mass transfer which, due to the extremely short nature of the CE, could be responsible for the secondaries being out of thermal equilibrium even after exiting the CE.  

The central star of \object{The Necklace} is a post-CE binary where the secondary shows clear evidence of being chemically polluted by AGB material from the primary \citep{miszalski13b,boffin14b}, proving that mass transfer can and does occur in these systems (and possibly in significant quantities - 0.03--0.35 M$_\odot$ in the case of \object{The Necklace}).  This mass transfer almost certainly occurred immediately prior to the CE, also being responsible for forming the polar jets seen in the nebula \cite[the dynamical ages of the jets are older than that of the central regions of the nebula, which is expected to be the remnant of the ejected CE,][]{corradi11full}.  Interestingly, amongst the other nebulae shown to play host to inflated secondaries, there is a second case where the polar regions of the nebula appear to have been ejected prior to the CE \cite[\object{Abell 63},][]{mitchell07a}.  Estimates of the mass transfer rate during the formation of the jets in these objects are of order 10$^{-5}$--10$^{-4}$ M$_\odot$\,yr$^{-1}$ \citep{tocknell14}, almost certainly sufficient to produce the inflation found in the secondaries.  It is worth noting that, while the other nebulae found to host inflated secondaries, including Hen~2-155, they have yet to be shown to present evidence of polar outflows formed prior to the CE phase. This may just be due to a lack of detailed morpho-kinematical study\footnote{This may likely be the case for Hen~2-155 given the presence of pairs geometrically opposed knots.  Further kinematical study will be essential not only in constraining the morphology of Hen~2-155 but also the nature of these ``jet'' candidates.} or simply because these outflows are generally extremely light and may since have simply dissipated into the surroundings and are no longer visible.

In this study, both Hen~2-155 and Hen~2-161 are found to display sub-solar, non-type \textsc{i} abundances (Hen~2-161 could be considered a marginal Type \textsc{i}) and higher than average \emph{adfs}, both of which have possible links to binarity.  Non-type \textsc{i} abundances can arise naturally from low-mass single star progenitors or high mass progenitors where a binary interaction has cut short the chemical evolution of the progenitor \citep{demarco09} - given that both nebulae display type \textsc{i}-like morphologies, this may be an indication that the latter scenario has played a role (however, this cannot be confirmed with the current observations).  Perhaps a stronger link to binarity is the high \emph{adf}, where recent work has shown that the PNe with the highest known \emph{adfs} all host a central binary star \citep[Jones et al.\ in prep; and references therein]{corradi15}.  High \emph{adfs} also strengthen the link between PNe with binary central stars and other binary phenomena, such as novae, which are also known to display extremely high \emph{adfs} \citep{wesson03,wesson08b}.

We strongly encourage the detailed study of other post-CE CSPN and their host nebulae (as well as the continued search for new systems) in order to improve the statistics and to establish the parameters space for future modelling and study of the CE and its effect on PN formation and evolution.  In the immediate future, this work should focus on radial velocity studies and modelling of eclipsing systems for which high-quality photometric monitoring has already been performed \citep[e.g.\ \object{M3~16}, \object{H~2-29} and \object{M~2-19};][]{miszalski08}, as eclipsing systems offer the best prospect for well-constrained parameters.  For non-eclipsing systems, the greatest uncertainty often arises from the a wide-range of inclinations offering reasonable fits to the data - in this cases, the now well established link between nebular inclination and binary inclination may be used to successfully break the degeneracy.  This, however, requires detailed spatio-kinematical modelling of the host nebulae \citep[e.g.][]{tyndall12,huckvale13} in order to determine the inclination of the nebular symmetry axis which can then be assumed to be roughly perpendicular to binary orbital plane \citep{jones12,jones14b,nordhaus06}.

\begin{acknowledgements}
We would like to thank Klaus Werner for his assistance in the interpretation of the stellar spectroscopy, and Tom Marsh for the use of his \textsc{molly} and \textsc{pamela} software packages. DJ would like to thank Jorge Garc\'ia-Rojas for useful discussions regarding the interpretation of the nebular spectroscopy.

This paper is based on observations made with ESO Telescopes at the La Silla Paranal Observatory under programme IDs 088.D-0573, 090.D-0435, 091.D-0475, 092.D-0449 \& 093.D-0038. This paper includes observations made with the Southern African Large Telescope (SALT) under programme 2012-1-RSA-009, and with the 1m telescope of the South African Astronomical Observatory (SAAO).  Based on observations made with the NASA/ESA Hubble Space Telescope, and obtained from the Hubble Legacy Archive, which is a collaboration between the Space Telescope Science Institute (STScI/NASA), the Space Telescope European Coordinating Facility (ST-ECF/ESA) and the Canadian Astronomy Data Centre (CADC/NRC/CSA). This research made use of NASA's Astrophysics Data System; the SIMBAD database, operated at CDS, Strasbourg, France; the VizieR catalogue access tool, CDS, Strasbourg, France.

This research has been supported by the Spanish Ministry of Economy and Competitiveness (MINECO) under grants AYA2012--38700 and AYA2012-35330. PRG acknowledges support from the MINECO under the Ram\'on y Cajal programme (RYC--2010--05762).

Finally, we wish to thank the anonymous referee for their comments which have improved the detail of this manuscript.
\end{acknowledgements}

\bibliographystyle{aa}
\bibliography{literature}

\onllongtab{2}{
\begin{longtable}{p{3cm} ccc}
\caption{\label{tab:155phot} A log of the \HBc{} photometric observations and measurements of the central star of Hen~2-155.  Data also available in machine-readable format via anonymous ftp from CDS.}\\
\hline\hline
Julian Date & Exposure time & \HBc\ magnitude & Uncertainty on \\
& (s)&& \HBc\ magnitude\\
\hline
\endfirsthead
\caption{continued.}\\
\hline\hline
Julian Date & Exposure time & \HBc\ magnitude & Uncertainty on \\
& (s)&& \HBc\ magnitude\\\hline
\endhead
\hline
\endfoot
2455985.7989840 & 90 &  15.0020 & 0.0304 \\ 
2455985.8008033 & 150 &  15.0008 & 0.0211 \\ 
2455985.8029363 & 150 &  15.0086 & 0.0213 \\ 
2455985.8050782 & 150 &  15.0051 & 0.0211 \\ 
2455985.8517651 & 150 &  14.8521 & 0.0193 \\ 
2455985.8538980 & 150 &  14.8548 & 0.0192 \\ 
2455985.8560267 & 150 &  14.8768 & 0.0197 \\ 
2455985.8928693 & 150 &  14.9180 & 0.0198 \\ 
2455985.8950082 & 150 &  14.9041 & 0.0204 \\ 
2455985.8971377 & 150 &  14.9504 & 0.0211 \\ 
2455985.8994509 & 150 &  14.9460 & 0.0226 \\ 
2455985.9015940 & 150 &  14.9688 & 0.0251 \\ 
2455985.9038309 & 150 &  14.9632 & 0.0293 \\ 
2455986.7685927 & 150 &  14.8737 & 0.0218 \\ 
2455986.7707301 & 150 &  14.8869 & 0.0207 \\ 
2455986.7728711 & 150 &  14.8874 & 0.0210 \\ 
2455986.7988129 & 150 &  14.9825 & 0.0204 \\ 
2455986.8009496 & 150 &  14.9838 & 0.0208 \\ 
2455986.8030796 & 150 &  14.9842 & 0.0204 \\ 
2455986.8424549 & 150 &  15.0068 & 0.0202 \\ 
2455986.8445840 & 150 &  15.0109 & 0.0203 \\ 
2455986.8467141 & 150 &  14.9983 & 0.0202 \\ 
2455986.8633496 & 150 &  14.9380 & 0.0195 \\ 
2455986.8654867 & 150 &  14.9186 & 0.0195 \\ 
2455986.8676278 & 150 &  14.9227 & 0.0195 \\ 
2455986.8989045 & 240 &  14.9124 & 0.0164 \\ 
2455986.9020726 & 240 &  14.9053 & 0.0190 \\ 
2455986.9052437 & 240 &  14.8986 & 0.0239 \\ 
2455986.9084149 & 240 &  14.8684 & 0.0315 \\ 
2455987.7737781 & 240 &  14.8719 & 0.0146 \\ 
2455987.7769572 & 240 &  14.8459 & 0.0146 \\ 
2455987.7801395 & 240 &  14.8638 & 0.0146 \\ 
2455987.8287887 & 240 &  14.9516 & 0.0144 \\ 
2455987.8319682 & 240 &  14.9647 & 0.0144 \\ 
2455987.8351401 & 240 &  14.9805 & 0.0145 \\ 
2455987.8557959 & 240 &  15.0370 & 0.0147 \\ 
2455987.8589712 & 240 &  15.1094 & 0.0151 \\ 
2455987.8621534 & 240 &  15.1781 & 0.0153 \\ 
2455987.8739850 & 240 &  15.0048 & 0.0145 \\ 
2455987.8771649 & 240 &  15.0116 & 0.0145 \\ 
2455987.8803363 & 240 &  14.9912 & 0.0145 \\ 
2455987.8835078 & 240 &  14.9943 & 0.0145 \\ 
2455987.8866787 & 240 &  14.9822 & 0.0145 \\ 
2455987.8898617 & 240 &  14.9740 & 0.0147 \\ 
2455988.7545551 & 240 &  15.1696 & 0.0170 \\ 
2455988.7577319 & 240 &  15.1223 & 0.0167 \\ 
2455988.7609034 & 240 &  15.0451 & 0.0159 \\ 
2455988.7640752 & 240 &  15.0063 & 0.0162 \\ 
2455988.7672461 & 240 &  14.9884 & 0.0161 \\ 
2455988.7704180 & 240 &  14.9918 & 0.0161 \\ 
2455988.8775318 & 240 &  14.9880 & 0.0145 \\ 
2455988.8807068 & 240 &  14.9985 & 0.0143 \\ 
2455989.8336887 & 240 &  14.9151 & 0.0149 \\ 
2455989.8368668 & 240 &  14.9035 & 0.0149 \\ 
2455989.8400383 & 240 &  14.8980 & 0.0147 \\ 
2455989.8693272 & 240 &  14.9338 & 0.0145 \\ 
2455989.8725036 & 240 &  14.8726 & 0.0146 \\ 
2455989.8885252 & 240 &  14.8834 & 0.0150 \\ 
2455989.8917048 & 240 &  14.8988 & 0.0152 \\ 
2456712.7485850 & 240 &  14.9415 & 0.0183 \\ 
2456712.7517599 & 240 &  14.9325 & 0.0181 \\ 
2456712.7549312 & 240 &  14.9680 & 0.0176 \\ 
2456712.7581045 & 240 &  14.9799 & 0.0180 \\ 
2456712.7612776 & 240 &  14.9929 & 0.0173 \\ 
2456712.7644514 & 240 &  14.9975 & 0.0174 \\ 
2456712.7676246 & 240 &  15.0094 & 0.0172 \\ 
2456712.7707959 & 240 &  15.0151 & 0.0173 \\ 
2456712.7739676 & 240 &  15.0231 & 0.0173 \\ 
2456712.7771418 & 240 &  15.0820 & 0.0175 \\ 
2456712.7803185 & 240 &  15.1568 & 0.0179 \\ 
2456712.7834991 & 240 &  15.1819 & 0.0183 \\ 
2456712.7866744 & 240 &  15.1511 & 0.0178 \\ 
2456712.7898470 & 240 &  15.0709 & 0.0174 \\ 
2456712.7930182 & 240 &  14.9734 & 0.0170 \\ 
2456712.7961942 & 240 &  14.9903 & 0.0171 \\ 
2456712.7993699 & 240 &  14.9847 & 0.0171 \\ 
2456712.8025435 & 240 &  14.9925 & 0.0169 \\ 
2456712.8057199 & 240 &  14.9825 & 0.0170 \\ 
2456712.8088957 & 240 &  14.9828 & 0.0168 \\ 
2456712.8120692 & 240 &  14.9572 & 0.0168 \\ 
2456712.8152405 & 240 &  14.9514 & 0.0166 \\ 
2456712.8184117 & 240 &  14.9383 & 0.0165 \\ 
2456712.8215838 & 240 &  14.9182 & 0.0166 \\ 
2456712.8247545 & 240 &  14.9205 & 0.0165 \\ 
2456712.8279307 & 240 &  14.9028 & 0.0165 \\ 
2456712.8311070 & 240 &  14.8989 & 0.0166 \\ 
2456712.8342797 & 240 &  14.8942 & 0.0164 \\ 
2456712.8406225 & 240 &  14.8618 & 0.0164 \\ 
2456712.8437979 & 240 &  14.8639 & 0.0161 \\ 
2456712.8469705 & 240 &  14.8600 & 0.0160 \\ 
2456712.8501450 & 240 &  14.8620 & 0.0162 \\ 
2456712.8533192 & 240 &  14.8933 & 0.0162 \\ 
2456712.8564960 & 240 &  14.9120 & 0.0165 \\ 
2456712.8596705 & 240 &  14.8935 & 0.0167 \\ 
2456712.8628436 & 240 &  14.8747 & 0.0162 \\ 
2456712.8660158 & 240 &  14.8529 & 0.0164 \\ 
2456712.8691906 & 240 &  14.8621 & 0.0162 \\ 
2456712.8723667 & 240 &  14.8753 & 0.0162 \\ 
2456712.8755420 & 240 &  14.8705 & 0.0164 \\ 
2456712.8787159 & 240 &  14.8787 & 0.0164 \\ 
2456712.8818900 & 240 &  14.8844 & 0.0166 \\ 
2456712.8850649 & 240 &  14.9206 & 0.0167 \\ 
2456712.8882415 & 240 &  14.9155 & 0.0174 \\ 
2456712.8914157 & 240 &  14.9155 & 0.0178 \\ 
2456712.8945906 & 240 &  14.9171 & 0.0192 \\ 
2456712.8977621 & 240 &  14.9348 & 0.0214 \\ 
2456713.7694692 & 240 &  14.8799 & 0.0166 \\ 
2456713.7821596 & 240 &  14.9281 & 0.0161 \\ 
2456713.8051735 & 240 &  14.9982 & 0.0167 \\ 
2456713.8197508 & 240 &  15.1681 & 0.0180 \\ 
2456713.8229228 & 240 &  15.1603 & 0.0178 \\ 
2456713.8260968 & 240 &  15.1091 & 0.0173 \\ 
2456713.8292670 & 240 &  15.0124 & 0.0173 \\ 
2456713.8324438 & 240 &  14.9939 & 0.0172 \\ 
2456713.8620228 & 240 &  14.9088 & 0.0163 \\ 
2456713.8801730 & 240 &  14.8479 & 0.0157 \\ 
2456714.8685630 & 240 &  14.9932 & 0.0162 \\ 
2456715.7739455 & 240 &  14.9621 & 0.0155 \\ 
2456715.8354988 & 240 &  14.8502 & 0.0155 \\ 
\end{longtable}
}

\onllongtab{3}{
\begin{longtable}{p{3cm} ccc}
\caption{\label{tab:161phot} A log of the $I$-band photometric observations and measurements of the central star of Hen~2-161.  Observations marked with an asterisk are from the SAAO 1.0-m telescope taken with a Cousins $I$-band filter, all other observations were acquired with NTT-EFOSC2 and a Gunn $i$ filter (see text for further details).  Data also available in machine-readable format via anonymous ftp from CDS.}\\
\hline\hline
Julian Date & Exposure time & $I$-band magnitude & Uncertainty on \\
& (s)&& $I$-band magnitude\\
\hline
\endfirsthead
\caption{continued.}\\
\hline\hline
Julian Date & Exposure time & $I$-band magnitude & Uncertainty on \\
& (s)&& $I$-band magnitude\\\hline
\endhead
\hline
\endfoot
2455987.7904517 & 45 &  15.2006 & 0.0120 \\ 
2455987.7913786 & 45 &  15.1932 & 0.0121 \\ 
2455987.7923043 & 45 &  15.2113 & 0.0120 \\ 
2455987.8456948 & 45 &  15.1422 & 0.0117 \\ 
2455987.8466095 & 45 &  15.1478 & 0.0116 \\ 
2455987.8475239 & 45 &  15.1570 & 0.0116 \\ 
2455987.8698281 & 45 &  15.1316 & 0.0111 \\ 
2455987.8707530 & 45 &  15.1248 & 0.0111 \\ 
2455987.8716673 & 45 &  15.1235 & 0.0113 \\ 
2455988.7951585 & 45 &  15.1798 & 0.0124 \\ 
2455988.7960771 & 45 &  15.1748 & 0.0124 \\ 
2455988.7969902 & 45 &  15.1782 & 0.0126 \\ 
2455988.7979061 & 45 &  15.1776 & 0.0129 \\ 
2455988.7988199 & 45 &  15.1794 & 0.0125 \\ 
2455988.8284592 & 45 &  15.1464 & 0.0126 \\ 
2455988.8293755 & 45 &  15.1623 & 0.0116 \\ 
2455988.8302899 & 45 &  15.1612 & 0.0120 \\ 
2455988.8312045 & 45 &  15.1596 & 0.0114 \\ 
2455988.8321303 & 45 &  15.1264 & 0.0132 \\ 
2455988.8330565 & 45 &  15.1494 & 0.0123 \\ 
2455988.8339706 & 45 &  15.1570 & 0.0122 \\ 
2455988.8734464 & 45 &  15.1181 & 0.0112 \\ 
2455988.8743642 & 45 &  15.1114 & 0.0120 \\ 
2455988.8752783 & 45 &  15.1063 & 0.0112 \\ 
2455988.9092162 & 45 &  15.0767 & 0.0197 \\ 
2455988.9101280 & 45 &  15.0795 & 0.0213 \\ 
2455988.9110539 & 45 &  15.0908 & 0.0236 \\ 
2455988.9120965 & 45 &  15.0711 & 0.0253 \\ 
2455988.9130097 & 45 &  15.0559 & 0.0293 \\ 
2455988.9139242 & 45 &  15.0799 & 0.0310 \\ 
2455989.7756987 & 60 &  15.1787 & 0.0110 \\ 
2455989.7767974 & 60 &  15.1788 & 0.0111 \\ 
2455989.7779935 & 60 &  15.1741 & 0.0115 \\ 
2455989.7790776 & 60 &  15.1792 & 0.0113 \\ 
2455989.8118550 & 60 &  15.1520 & 0.0111 \\ 
2455989.8129437 & 60 &  15.1667 & 0.0112 \\ 
2455989.8140312 & 60 &  15.1646 & 0.0114 \\ 
2455989.8151193 & 60 &  15.1451 & 0.0123 \\ 
2455989.8162188 & 60 &  15.1596 & 0.0115 \\ 
2455989.8173181 & 60 &  15.1532 & 0.0117 \\ 
2455989.8184179 & 60 &  15.1484 & 0.0115 \\ 
2455989.8195056 & 60 &  15.1611 & 0.0108 \\ 
2455989.8205937 & 60 &  15.1574 & 0.0113 \\ 
2455989.8216820 & 60 &  15.1506 & 0.0114 \\ 
2455989.8542936 & 60 &  15.1465 & 0.0097 \\ 
2455989.8553856 & 60 &  15.1463 & 0.0097 \\ 
2455989.8564737 & 60 &  15.1477 & 0.0095 \\ 
2455989.8575615 & 60 &  15.1357 & 0.0099 \\ 
2455989.8586609 & 60 &  15.1436 & 0.0096 \\ 
$\ast$ 2456020.4300300 & 90 &  15.0213 & 0.0239 \\ 
$\ast$ 2456020.4317400 & 90 &  14.9752 & 0.0243 \\ 
$\ast$ 2456020.4334400 & 90 &  14.9865 & 0.0245 \\ 
$\ast$ 2456020.4351400 & 90 &  14.9703 & 0.0245 \\ 
$\ast$ 2456020.4368300 & 90 &  15.0060 & 0.0237 \\ 
$\ast$ 2456020.4385300 & 90 &  15.0092 & 0.0241 \\ 
$\ast$ 2456020.4643600 & 90 &  15.0226 & 0.0227 \\ 
$\ast$ 2456020.4660600 & 90 &  15.0153 & 0.0236 \\ 
$\ast$ 2456020.4677700 & 90 &  15.0053 & 0.0238 \\ 
$\ast$ 2456020.4694800 & 90 &  14.9978 & 0.0236 \\ 
$\ast$ 2456020.4711700 & 90 &  15.0191 & 0.0236 \\ 
$\ast$ 2456020.4728700 & 90 &  14.9946 & 0.0233 \\ 
$\ast$ 2456020.5401500 & 90 &  14.9814 & 0.0227 \\ 
$\ast$ 2456020.5418500 & 90 &  15.0005 & 0.0222 \\ 
$\ast$ 2456020.5437800 & 90 &  14.9933 & 0.0242 \\ 
$\ast$ 2456020.5453100 & 90 &  15.0016 & 0.0240 \\ 
$\ast$ 2456020.5468400 & 90 &  14.9933 & 0.0242 \\ 
$\ast$ 2456020.5483700 & 90 &  14.9922 & 0.0243 \\ 
$\ast$ 2456020.5498800 & 90 &  14.9934 & 0.0245 \\ 
$\ast$ 2456020.5514200 & 90 &  15.0069 & 0.0239 \\ 
2456308.8388791 & 60 &  15.1144 & 0.0122 \\ 
2456308.8399697 & 60 &  15.0490 & 0.0131 \\ 
2456308.8410683 & 60 &  15.0941 & 0.0129 \\ 
2456712.9020590 & 60 &  15.0705 & 0.0138 \\ 
2456712.9031510 & 60 &  15.0727 & 0.0153 \\ 
2456712.9042402 & 60 &  15.0441 & 0.0162 \\ 
2456712.9064182 & 60 &  15.0333 & 0.0193 \\ 
2456712.9075069 & 60 &  15.1185 & 0.0211 \\ 
2456712.9085973 & 60 &  15.0293 & 0.0223 \\ 
2456713.7874744 & 30 &  15.1397 & 0.0132 \\ 
2456713.8696886 & 30 &  15.1325 & 0.0098 \\ 
2456713.8707783 & 30 &  15.1018 & 0.0101 \\ 
2456713.8718675 & 30 &  15.1009 & 0.0103 \\ 
2456713.8729605 & 30 &  15.0926 & 0.0137 \\ 
2456713.8736989 & 30 &  15.0973 & 0.0111 \\ 
2456713.8746154 & 30 &  15.0968 & 0.0131 \\ 
2456713.8753560 & 30 &  15.0826 & 0.0134 \\ 
2456713.9050826 & 30 &  15.0255 & 0.0170 \\ 
2456713.9058287 & 30 &  15.0434 & 0.0173 \\ 
2456713.9065719 & 30 &  15.0360 & 0.0183 \\ 
2456713.9073110 & 30 &  15.0348 & 0.0191 \\ 
2456713.9080531 & 30 &  15.0387 & 0.0196 \\ 
2456713.9087938 & 30 &  15.0381 & 0.0211 \\ 
2456713.9095347 & 30 &  15.0664 & 0.0221 \\ 
2456713.9102812 & 30 &  15.0268 & 0.0239 \\ 
2456713.9110239 & 30 &  15.0735 & 0.0246 \\ 
2456713.8366994 & 30 &  15.1993 & 0.0138 \\ 
2456713.8377933 & 30 &  15.0824 & 0.0128 \\ 
2456713.8388863 & 30 &  15.1242 & 0.0128 \\ 
2456715.7651854 & 30 &  15.1765 & 0.0139 \\ 
2456715.7659303 & 30 &  15.1752 & 0.0139 \\ 
2456715.7666715 & 30 &  15.1780 & 0.0141 \\ 
2456715.7674132 & 30 &  15.1861 & 0.0139 \\ 
2456715.7681527 & 30 &  15.1853 & 0.0139 \\ 
2456715.7688957 & 30 &  15.1679 & 0.0139 \\ 
2456715.7696373 & 30 &  15.1724 & 0.0139 \\ 
2456715.7703794 & 30 &  15.1717 & 0.0137 \\ 
2456715.7711246 & 30 &  15.1757 & 0.0137 \\ 
2456715.7718605 & 30 &  15.1818 & 0.0138 \\ 
2456715.7821279 & 30 &  15.1772 & 0.0142 \\ 
2456715.7828718 & 30 &  15.1800 & 0.0138 \\ 
2456715.7836141 & 30 &  15.1942 & 0.0138 \\ 
2456715.7843573 & 30 &  15.1833 & 0.0135 \\ 
2456715.7851024 & 30 &  15.1680 & 0.0142 \\ 
2456715.7997002 & 30 &  15.1677 & 0.0136 \\ 
2456715.8004471 & 30 &  15.1670 & 0.0138 \\ 
2456715.8011896 & 30 &  15.1870 & 0.0134 \\ 
2456715.8019356 & 30 &  15.1745 & 0.0136 \\ 
2456715.8026814 & 30 &  15.1713 & 0.0134 \\ 
2456715.8163368 & 30 &  15.1464 & 0.0140 \\ 
2456715.8170778 & 30 &  15.1751 & 0.0135 \\ 
2456715.8178196 & 30 &  15.1611 & 0.0137 \\ 
2456715.8185615 & 30 &  15.1565 & 0.0140 \\ 
2456715.8193035 & 30 &  15.1515 & 0.0138 \\ 
2456715.8275854 & 30 &  15.1554 & 0.0134 \\ 
2456715.8283297 & 30 &  15.1479 & 0.0132 \\ 
2456715.8290752 & 30 &  15.1623 & 0.0133 \\ 
2456715.8298200 & 30 &  15.1427 & 0.0135 \\ 
2456715.8305639 & 30 &  15.1541 & 0.0134 \\ 
2456715.8397072 & 30 &  15.1192 & 0.0143 \\ 
2456715.8404538 & 30 &  15.1482 & 0.0155 \\ 
2456715.8411984 & 30 &  15.0934 & 0.0155 \\ 
2456715.8419406 & 30 &  15.1156 & 0.0141 \\ 
2456715.8426877 & 30 &  15.1134 & 0.0145 \\ 
2456715.8553575 & 30 &  15.1016 & 0.0138 \\ 
2456715.8561049 & 30 &  15.1108 & 0.0140 \\ 
2456715.8568454 & 30 &  15.1145 & 0.0139 \\ 
2456715.8575932 & 30 &  15.1190 & 0.0139 \\ 
2456715.8583346 & 30 &  15.1140 & 0.0139 \\ 
2456715.8657918 & 30 &  15.0939 & 0.0147 \\ 
2456715.8665369 & 30 &  15.0773 & 0.0140 \\ 
2456715.8672778 & 30 &  15.0757 & 0.0143 \\ 
2456715.8680232 & 30 &  15.0826 & 0.0139 \\ 
2456715.8687678 & 30 &  15.1047 & 0.0136 \\ 
2456715.8783854 & 30 &  15.0929 & 0.0130 \\ 
2456715.8791289 & 30 &  15.0793 & 0.0133 \\ 
2456715.8798703 & 30 &  15.0834 & 0.0130 \\ 
2456715.8806149 & 30 &  15.1066 & 0.0131 \\ 
2456715.8813587 & 30 &  15.1092 & 0.0125 \\ 
2456715.8821031 & 30 &  15.1071 & 0.0124 \\ 
2456715.8828448 & 30 &  15.1102 & 0.0128 \\ 
2456715.8835893 & 30 &  15.1065 & 0.0126 \\ 
2456715.8843321 & 30 &  15.1066 & 0.0124 \\ 
2456715.8850763 & 30 &  15.1189 & 0.0122 \\ 
2456715.8858229 & 30 &  15.1100 & 0.0121 \\ 
2456715.8865687 & 30 &  15.1047 & 0.0123 \\ 
2456715.8873128 & 30 &  15.1171 & 0.0122 \\ 
2456715.8880534 & 30 &  15.1035 & 0.0122 \\ 
2456715.8887966 & 30 &  15.0976 & 0.0124 \\ 
2456715.8895430 & 30 &  15.0848 & 0.0124 \\ 
2456715.8902869 & 30 &  15.1029 & 0.0124 \\ 
2456715.8910275 & 30 &  15.0973 & 0.0123 \\ 
2456715.8917722 & 30 &  15.0993 & 0.0123 \\ 
2456715.8925178 & 30 &  15.1040 & 0.0122 \\ 
2456715.9027146 & 30 &  15.0835 & 0.0153 \\ 
2456715.9034564 & 30 &  15.0735 & 0.0163 \\ 
2456715.9041989 & 30 &  15.0818 & 0.0175 \\ 
2456715.9049431 & 30 &  15.0771 & 0.0177 \\ 
2456715.9056870 & 30 &  15.0840 & 0.0185 \\ 
2456715.9064324 & 30 &  15.0778 & 0.0197 \\ 
2456715.9071736 & 30 &  15.0769 & 0.0213 \\ 
2456715.9079143 & 30 &  15.0791 & 0.0226 \\ 
2456715.9086595 & 30 &  15.0533 & 0.0239 \\ 
2456715.9094049 & 30 &  15.0890 & 0.0262 \\ 
2456715.9101482 & 30 &  15.0866 & 0.0282 \\ 
2456715.9108930 & 30 &  15.0884 & 0.0302 \\ 
2456715.9116378 & 30 &  15.1014 & 0.0315 \\ 
2456715.9123814 & 30 &  15.0818 & 0.0346 \\ 
2456715.9131226 & 30 &  15.0021 & 0.0391 \\ 
2456715.9138636 & 30 &  15.1047 & 0.0445 \\ 
2456715.9146030 & 30 &  15.1301 & 0.0470 \\ 
2456715.9153486 & 30 &  15.1707 & 0.0530 \\ 
\end{longtable}
}

\onllongtab{4}{
\begin{longtable}{p{3cm} cc}
\caption{\label{tab:155rv} A log of the radial velocity measurements of the central star of Hen~2-155.  Data also available in machine-readable format via anonymous ftp from CDS.}\\
\hline\hline
Julian Date & Radial Velocity & Uncertainty on radial velocity \\
&(\kms{}) & (\kms{}) \\
\hline
\endfirsthead
\caption{continued.}\\
\hline\hline
Julian Date & Radial Velocity & Uncertainty on radial velocity \\
&(\kms{}) & (\kms{}) \\
\hline
\endhead
\hline
\endfoot
2456371.7604575 & -27.9696 & 2.8980 \\ 
2456371.7720789 & -50.7702 & 3.4633 \\ 
2456371.7837002 & -63.7826 & 3.3738 \\ 
2456371.7953216 & -53.4388 & 3.7784 \\ 
2456371.8069431 & -29.0446 & 3.2967 \\ 
2456371.8185528 & -8.8539 & 2.5652 \\ 
2456371.8301742 & 23.8644 & 2.7731 \\ 
2456371.8417956 & 50.8250 & 2.5643 \\ 
2456371.8534171 & 56.5280 & 2.4353 \\ 
2456371.8650384 & 59.9633 & 3.1728 \\ 
2456371.8766599 & 35.6776 & 3.4437 \\ 
2456371.8882813 & 15.7247 & 2.7868 \\ 
2456371.9016389 & -18.8627 & 2.6239 \\ 
\end{longtable}
}

\onllongtab{5}{
\begin{longtable}{p{3cm} cc}
\caption{\label{tab:155rvCiv} A log of the radial velocity measurements of the C~\textsc{iv} lines of Hen~2-155.  Data also available in machine-readable format via anonymous ftp from CDS.}\\
\hline\hline
Julian Date & Radial Velocity & Uncertainty on radial velocity \\
&(\kms{}) & (\kms{}) \\
\hline
\endfirsthead
\caption{continued.}\\
\hline\hline
Julian Date & Radial Velocity & Uncertainty on radial velocity \\
&(\kms{}) & (\kms{}) \\
\hline
\endhead
\hline
\endfoot
2456371.7604575 & -20.5399 & 2.7514 \\ 
2456371.7720789 & -22.5216 & 2.6793 \\ 
2456371.7837002 & -35.5964 & 3.2085 \\ 
2456371.7953216 & -38.9358 & 2.9990 \\ 
2456371.8069431 & -24.4214 & 2.5088 \\ 
2456371.8185528 & 2.7566 & 5.1585  \\ 
2456371.8301742 & 7.9988 & 3.1546  \\ 
2456371.8417956 & 19.2400 & 3.2090  \\ 
2456371.8534171 & 31.3309 & 2.4482  \\ 
2456371.8650384 & 36.1928 & 3.1091  \\ 
2456371.8766599 & 21.5965 & 2.6399  \\ 
2456371.8882813 & 19.7743 & 2.4719  \\ 
2456371.9016389 & 6.1171 & 2.5737  \\
\end{longtable}
}

\onllongtab{6}{
\begin{longtable}{llrlrlrlrl}
\caption{\label{tab:fluxes} Observed, $F(\lambda)$, and dereddened, $I(\lambda)$, nebular emission line fluxes from Hen~2-155 and Hen~2-161}\\
\hline\hline
  \multicolumn{2}{c}{} & \multicolumn{4}{c}{Hen2-155} &\multicolumn{4}{c}{Hen2-161}\\ \hline
    $ \lambda $ (\AA{}) & Ion & \multicolumn{2}{c}{$F \left( \lambda \right) $ }& \multicolumn{2}{c}{$I \left( \lambda \right) $}& \multicolumn{2}{c}{$F \left( \lambda \right) $} & \multicolumn{2}{c}{$I \left( \lambda \right) $} \\ \hline
  \endfirsthead
  \caption{continued.}\\
  \hline \hline
  \multicolumn{2}{c}{} & \multicolumn{4}{c}{Hen2-155} &\multicolumn{4}{c}{Hen2-161}\\ \hline
    $ \lambda $ (\AA{}) & Ion & \multicolumn{2}{c}{$F \left( \lambda \right) $ }& \multicolumn{2}{c}{$I \left( \lambda \right) $}& \multicolumn{2}{c}{$F \left( \lambda \right) $} & \multicolumn{2}{c}{$I \left( \lambda \right) $} \\ \hline
\endhead
\endfoot
3664.68 & H~{\sc i}       &    0.28& $\pm$   0.05&   0.43& $^{  +0.07}_{  -0.08}$ & - & - & - & -\\
 3666.10 & H~{\sc i}       &    0.26& $\pm$   0.05&   0.38& $^{  +0.07}_{  -0.08}$ &    0.41& $\pm$   0.08&   0.85& $^{  +0.14}_{  -0.17}$ \\
 3667.68 & H~{\sc i}       & \multicolumn{2}{c}{-} & \multicolumn{2}{c}{-}&    0.20& $\pm$   0.05&   0.26& $^{  +0.05}_{  -0.06}$ \\
  3669.46 & H~{\sc i}       &    0.26& $\pm$   0.08&   0.42& $^{  +0.09}_{  -0.11}$ & - & - & - & - \\
 3709.62 & Ne~{\sc ii}     & \multicolumn{2}{c}{-} & \multicolumn{2}{c}{-}&    0.27& $\pm$   0.05&   0.53& $^{  +0.09}_{  -0.11}$\\
 3711.97 & H~{\sc i}       &    0.64& $\pm$   0.07&   1.00& $\pm$   0.11 &    0.84& $\pm$   0.08&   1.72& $\pm$   0.15 \\
 3713.08 & Ne~{\sc ii}     &    0.35& $\pm$   0.08&   0.48& $^{  +0.09}_{  -0.11}$ & - & - & - & -\\
 3721.94 & H~{\sc i}       &    1.50& $\pm$   0.14&   2.33& $\pm$   0.21 &    1.07& $\pm$   0.13&   2.20& $\pm$   0.27 \\
 3726.03 & [O~{\sc ii}]    &   15.55& $\pm$   0.15&  24.07& $\pm$   0.24 &    8.81& $\pm$   0.10&  18.02& $\pm$   0.23 \\
 3728.82 & [O~{\sc ii}]    &   11.60& $\pm$   0.16&  17.94& $\pm$   0.25  &    5.70& $\pm$   0.08&  11.64& $\pm$   0.18\\
 3734.37 & H~{\sc i}       &    1.36& $\pm$   0.09&   2.10& $\pm$   0.13 &    0.97& $\pm$   0.07&   1.98& $\pm$   0.14\\
 3750.15 & H~{\sc i}       &    1.92& $\pm$   0.04&   2.95& $\pm$   0.06 &    1.28& $\pm$   0.04&   2.58& $\pm$   0.08\\
 3770.63 & H~{\sc i}       &    2.20& $\pm$   0.07&   3.36& $\pm$   0.11 &    1.52& $\pm$   0.09&   3.03& $\pm$   0.18\\
 3797.90 & H~{\sc i}       &    2.90& $\pm$   0.08&   4.39& $\pm$   0.12 &    1.97& $\pm$   0.11&   3.89& $\pm$   0.22\\
 3856.13 & O~{\sc ii}      & \multicolumn{2}{c}{-} & \multicolumn{2}{c}{-}&    0.17& $\pm$   0.05&   0.25& $^{  +0.05}_{  -0.07}$\\
 3868.75 & [Ne~{\sc iii}]  &   56.77& $\pm$   0.08&  84.06& $\pm$   0.21 &   15.38& $\pm$   0.13&  29.23& $\pm$   0.28 \\
 3888.65 & He~{\sc i}      &    9.10& $\pm$   0.05&  13.38& $\pm$   0.08 &    7.47& $\pm$   0.06&  14.05& $\pm$   0.14 \\
 3889.05 & H~{\sc i}       &    5.36& $\pm$   0.06&   7.88& $\pm$   0.08  &    4.35& $\pm$   0.07&   8.17& $\pm$   0.14 \\
  3964.73 & He~{\sc i}      & \multicolumn{2}{c}{-} & \multicolumn{2}{c}{-}&    0.70& $\pm$   0.15&   1.16& $^{  +0.20}_{  -0.25}$\\
 3967.46 & [Ne~{\sc iii}]  &   17.56& $\pm$   0.27&  25.13& $\pm$   0.39 &    4.48& $\pm$   0.15&   8.06& $\pm$   0.27\\
 3970.07 & H~{\sc i}       &    9.36& $\pm$   0.26&  13.38& $\pm$   0.38 &    6.92& $\pm$   0.13&  12.43& $\pm$   0.25\\
 4026.21 & He~{\sc i}      &    1.42& $\pm$   0.03&   1.99& $\pm$   0.04  &    1.48& $\pm$   0.04&   2.57& $\pm$   0.08\\
 4068.60 & [S~{\sc ii}]    &    0.39& $\pm$   0.02&   0.54& $\pm$   0.02  &    0.28& $\pm$   0.03&   0.47& $\pm$   0.04\\
 4069.62 & O~{\sc ii}      &    0.26& $\pm$   0.01&   0.37& $\pm$   0.02  &    0.31& $\pm$   0.03&   0.53& $\pm$   0.04\\
 4072.16 & O~{\sc ii}      &    0.24& $\pm$   0.01&   0.32& $\pm$   0.02 &    0.38& $\pm$   0.03&   0.63& $\pm$   0.05 \\
 4075.86 & O~{\sc ii}      &    0.30& $\pm$   0.02&   0.42& $\pm$   0.03  &    0.39& $\pm$   0.03&   0.65& $\pm$   0.06\\
 4076.35 & [S~{\sc ii}]    &    0.17& $\pm$   0.02&   0.23& $\pm$   0.03  & \multicolumn{2}{c}{-} & \multicolumn{2}{c}{-}\\
 4078.84 & O~{\sc ii}      &    0.06& $\pm$   0.02&   0.06& $^{  +0.01}_{  -0.02}$ & \multicolumn{2}{c}{-} & \multicolumn{2}{c}{-}\\
 4085.11 & O~{\sc ii}      &    0.07& $\pm$   0.02&   0.08& $\pm$   0.02  & \multicolumn{2}{c}{-} & \multicolumn{2}{c}{-}\\
  4087.15 & O~{\sc ii}      & \multicolumn{2}{c}{-} & \multicolumn{2}{c}{-}&    0.19& $\pm$   0.03&   0.31& $^{  +0.05}_{  -0.06}$\\
 4089.29 & O~{\sc ii}      &    0.15& $\pm$   0.02&   0.20& $\pm$   0.03  &    0.41& $\pm$   0.03&   0.69& $\pm$   0.04\\
 4097.26 & O~{\sc ii}      &    0.79& $\pm$   0.06&   1.07& $\pm$   0.08  &    0.36& $\pm$   0.04&   0.59& $\pm$   0.07\\
 4101.74 & H~{\sc i}       &   17.93& $\pm$   0.06&  24.44& $\pm$   0.09 &   15.00& $\pm$   0.04&  24.88& $\pm$   0.13\\
 4120.28 & O~{\sc ii}      &    0.13& $\pm$   0.02&   0.17& $\pm$   0.03 & \multicolumn{2}{c}{-} & \multicolumn{2}{c}{-} \\
 4121.46 & O~{\sc ii}      &    0.17& $\pm$   0.05&   0.23& $^{  +0.04}_{  -0.06}$ & \multicolumn{2}{c}{-} & \multicolumn{2}{c}{-}\\
 4132.80 & O~{\sc ii}      &    0.08& $\pm$   0.02&   0.09& $\pm$   0.02 & \multicolumn{2}{c}{-} & \multicolumn{2}{c}{-}\\
 4153.30 & O~{\sc ii}      &    0.11& $\pm$   0.02&   0.14& $^{  +0.02}_{  -0.03}$ &    0.22& $\pm$   0.03&   0.35& $\pm$   0.05  \\
 4156.53 & O~{\sc ii}      & \multicolumn{2}{c}{-} & \multicolumn{2}{c}{-}&    0.13& $\pm$   0.03&   0.18& $^{  +0.03}_{  -0.04}$\\
 4186.90 & C~{\sc iii}     &    0.09& $\pm$   0.02&   0.12& $\pm$   0.02 & \multicolumn{2}{c}{-} & \multicolumn{2}{c}{-} \\
 4237.05 & N~{\sc ii}      & \multicolumn{2}{c}{-} & \multicolumn{2}{c}{-}&    0.10& $\pm$   0.02&   0.15& $\pm$   0.02\\
  4241.78 & N~{\sc ii}      & \multicolumn{2}{c}{-} & \multicolumn{2}{c}{-}&    0.10& $\pm$   0.02&   0.15& $^{  +0.02}_{  -0.03}$\\
 4267.15 & C~{\sc ii}      &    0.24& $\pm$   0.01&   0.30& $\pm$   0.01 &    1.56& $\pm$   0.03&   2.34& $\pm$   0.04 \\
 4273.10 & O~{\sc ii}      &    0.04& $\pm$   0.01&   0.03& $\pm$   0.01  & \multicolumn{2}{c}{-} & \multicolumn{2}{c}{-}\\
 4275.55 & O~{\sc ii}      &    0.04& $\pm$   0.01&   0.04& $\pm$   0.01  & \multicolumn{2}{c}{-} & \multicolumn{2}{c}{-}\\
 4276.75 & O~{\sc ii}      & \multicolumn{2}{c}{-} & \multicolumn{2}{c}{-}&    0.12& $\pm$   0.01&   0.18& $\pm$   0.02\\
 4275.99 & O~{\sc ii}      &    0.03& $\pm$   0.01&   0.03& $\pm$   0.01  & \multicolumn{2}{c}{-} & \multicolumn{2}{c}{-}\\
 4277.89 & O~{\sc ii}      & \multicolumn{2}{c}{-} & \multicolumn{2}{c}{-}&    0.10& $\pm$   0.01&   0.15& $\pm$   0.02\\
  4282.96 & O~{\sc ii}      & \multicolumn{2}{c}{-} & \multicolumn{2}{c}{-}&    0.07& $\pm$   0.01&   0.11& $\pm$ 0.02\\
   4285.69 & O~{\sc ii}      & \multicolumn{2}{c}{-} & \multicolumn{2}{c}{-}&    0.06& $\pm$   0.01&   0.09& $^{  +0.01}_{  -0.02}$\\
 4291.25 & O~{\sc ii}      &    0.03& $\pm$   0.01&   0.03& $\pm$   0.01  & \multicolumn{2}{c}{-} & \multicolumn{2}{c}{-}\\
 4294.78 & O~{\sc ii}      &    0.05& $\pm$   0.01&   0.06& $\pm$   0.01 &    0.09& $\pm$   0.01&   0.13& $\pm$   0.02  \\
 4303.61 & O~{\sc ii}      &    0.11& $\pm$   0.01&   0.13& $\pm$   0.01 &    0.17& $\pm$   0.03&   0.25& $\pm$   0.04 \\
 4317.14 & O~{\sc ii}      & \multicolumn{2}{c}{-} & \multicolumn{2}{c}{-}&    0.16& $\pm$   0.02&   0.24& $\pm$   0.02\\
 4317.70 & O~{\sc ii}      & \multicolumn{2}{c}{-} & \multicolumn{2}{c}{-}&    0.05& $\pm$   0.02&   0.06& $^{  +0.01}_{  -0.02}$\\
 4319.63 & O~{\sc ii}      & \multicolumn{2}{c}{-} & \multicolumn{2}{c}{-}&    0.10& $\pm$   0.02&   0.15& $\pm$   0.02\\
 4340.47 & H~{\sc i}       &   34.68& $\pm$   0.08&  43.04& $\pm$   0.12 &   32.08& $\pm$   0.17&  45.67& $\pm$   0.29\\
 4345.55 & O~{\sc ii}      & \multicolumn{2}{c}{-} & \multicolumn{2}{c}{-}&    0.45& $\pm$   0.08&   0.63& $^{  +0.10}_{  -0.12}$\\
 4349.43 & O~{\sc ii}      &    0.12& $\pm$   0.02&   0.14& $^{  +0.02}_{  -0.03}$  &    0.29& $\pm$   0.05&   0.40& $^{  +0.07}_{  -0.08}$ \\
 4363.21 & [O~{\sc iii}]   &    5.67& $\pm$   0.02&   6.97& $\pm$   0.03 &    0.60& $\pm$   0.03&   0.84& $\pm$   0.04 \\
 4366.89 & O~{\sc ii}      &    0.09& $\pm$   0.02&   0.09& $\pm$   0.02  &    0.23& $\pm$   0.02&   0.32& $\pm$   0.03\\
 4379.11 & N~{\sc iii}     &    0.16& $\pm$   0.01&   0.20& $\pm$   0.02 &    0.09& $\pm$   0.03&   0.09& $^{  +0.02}_{  -0.03}$ \\
 4387.93 & He~{\sc i}      &    0.47& $\pm$   0.01&   0.57& $\pm$   0.01 &    0.54& $\pm$   0.03&   0.75& $\pm$   0.03 \\
 4416.97 & O~{\sc ii}      & \multicolumn{2}{c}{-} & \multicolumn{2}{c}{-}&    0.10& $\pm$   0.03&   0.12& $^{  +0.02}_{  -0.03}$\\
 4437.55 & He~{\sc i}      &    0.04& $\pm$   0.01&   0.03& $\pm$   0.01 & \multicolumn{2}{c}{-} & \multicolumn{2}{c}{-}\\
 4452.37 & O~{\sc ii}      &    0.04& $\pm$   0.01&   0.04& $\pm$   0.01  &    0.16& $\pm$   0.04&   0.18& $^{  +0.03}_{  -0.04}$\\
 4471.50 & He~{\sc i}      &    4.21& $\pm$   0.02&   4.96& $\pm$   0.03  &    4.91& $\pm$   0.06&   6.41& $\pm$   0.08\\
 4530.41 & N~{\sc ii}      & \multicolumn{2}{c}{-} & \multicolumn{2}{c}{-}&    0.06& $\pm$   0.02&   0.07& $^{  +0.01}_{  -0.02}$\\
 4552.53 & N~{\sc ii}      &    0.05& $\pm$   0.02&   0.05& $\pm$   0.01 & \multicolumn{2}{c}{-} & \multicolumn{2}{c}{-} \\
 4601.48 & N~{\sc ii}      &    0.07& $\pm$   0.01&   0.08& $\pm$   0.01 & \multicolumn{2}{c}{-} & \multicolumn{2}{c}{-} \\
 4609.44 & O~{\sc ii}      &    0.03& $\pm$   0.01&   0.03& $\pm$   0.01 &    0.13& $\pm$   0.02&   0.15& $\pm$   0.02 \\
 4610.20 & O~{\sc ii}      &    0.06& $\pm$   0.01&   0.06& $\pm$   0.01  & \multicolumn{2}{c}{-} & \multicolumn{2}{c}{-}\\
 4630.54 & N~{\sc ii}      &    0.06& $\pm$   0.02&   0.06& $\pm$   0.01  &    0.20& $\pm$   0.03&   0.23& $\pm$   0.03\\
 4640.64 & N~{\sc iii}     &    0.77& $\pm$   0.02&   0.85& $\pm$   0.02  &    0.53& $\pm$   0.02&   0.61& $\pm$   0.02\\
 4647.42 & C~{\sc iii}     &    0.08& $\pm$   0.01&   0.09& $\pm$   0.01  & \multicolumn{2}{c}{-} & \multicolumn{2}{c}{-}\\
 4649.13 & O~{\sc ii}      &    0.37& $\pm$   0.01&   0.40& $\pm$   0.02  &    0.87& $\pm$   0.03&   1.00& $\pm$   0.04\\
 4650.25 & C~{\sc iii}     &    0.04& $\pm$   0.01&   0.03& $\pm$   0.01   & \multicolumn{2}{c}{-} & \multicolumn{2}{c}{-}\\
 4650.84 & O~{\sc ii}      &    0.11& $\pm$   0.01&   0.12& $\pm$   0.02 &    0.20& $\pm$   0.03&   0.23& $\pm$  0.04 \\
 4661.63 & O~{\sc ii}      &    0.14& $\pm$   0.01&   0.16& $\pm$   0.02 &    0.32& $\pm$   0.02&   0.36& $\pm$   0.02 \\
 4685.68 & He~{\sc ii}     &    7.45& $\pm$   0.03&   8.02& $\pm$   0.04 &    0.45& $\pm$   0.10&   0.46& $^{  +0.08}_{  -0.10}$ \\
 4696.35 & O~{\sc ii}      & \multicolumn{2}{c}{-} & \multicolumn{2}{c}{-}&    0.08& $\pm$   0.03&   0.07& $\pm$  0.02\\
 4711.37 & [Ar~{\sc iv}]   &    2.32& $\pm$   0.02&   2.47& $\pm$   0.03 &    0.34& $\pm$   0.04&   0.37& $\pm$   0.05 \\
 4713.17 & He~{\sc i}      &    0.38& $\pm$   0.02&   0.41& $\pm$   0.02 &    0.45& $\pm$   0.04&   0.50& $\pm$   0.05\\
 4714.17 & [Ne~{\sc iv}]   &    0.07& $\pm$   0.02&   0.06& $\pm$   0.01  & \multicolumn{2}{c}{-} & \multicolumn{2}{c}{-}\\
 4725.62 & [Ne~{\sc iv}]   & \multicolumn{2}{c}{-} & \multicolumn{2}{c}{-}&    0.07& $\pm$   0.02&   0.06& $\pm$ 0.01\\
 4740.17 & [Ar~{\sc iv}]   &    1.85& $\pm$   0.03&   1.95& $\pm$   0.03 &    0.22& $\pm$   0.05&   0.21& $^{  +0.04}_{  -0.05}$  \\
 4861.33 & H~{\sc i}       &  100.00& $\pm$   0.13& 100.00& $\pm$0.15 &  100.00& $\pm$   0.29& 100.00& $\pm$ 0.30\\
 4906.83 & O~{\sc ii}      & \multicolumn{2}{c}{-} & \multicolumn{2}{c}{-}&    0.12& $\pm$   0.02&   0.12& $\pm$   0.02\\
 4921.93 & He~{\sc i}      &    1.42& $\pm$   0.04&   1.38& $\pm$   0.04  &    1.70& $\pm$   0.03&   1.63& $\pm$   0.03\\
  4924.53 & O~{\sc ii}      & \multicolumn{2}{c}{-} & \multicolumn{2}{c}{-}&    0.16& $\pm$   0.03&   0.14& $^{  +0.02}_{  -0.03}$\\
 4958.91 & [O~{\sc iii}]   &  238.20& $\pm$   2.22& 228.68& $\pm$   2.15 &  101.91& $\pm$   0.32&  95.33& $\pm$   0.40 \\
 5006.84 & [O~{\sc iii}]   & \multicolumn{2}{c}{-} & \multicolumn{2}{c}{-}&  303.14& $\pm$   2.03& 274.44& $\pm$   1.99\\
 5015.68 & He~{\sc i}      &    4.15& $\pm$   0.84&   3.63& $^{  +0.62}_{  -0.75}$& \multicolumn{2}{c}{-} & \multicolumn{2}{c}{-}  \\
 5754.60 & [N~{\sc ii}]    & \multicolumn{2}{c}{-} & \multicolumn{2}{c}{-}&    1.08& $\pm$   0.10&   0.63& $\pm$   0.06\\
 5875.66 & He~{\sc i}      & \multicolumn{2}{c}{-} & \multicolumn{2}{c}{-}&   34.56& $\pm$   0.25&  19.00& $\pm$   0.14\\
 5931.78 & N~{\sc ii}      &    0.23& $\pm$   0.07&   0.12& $\pm$   0.03  &    0.33& $\pm$   0.09&   0.15& $\pm$ 0.03\\
 5941.65 & N~{\sc ii}      &    0.27& $\pm$   0.07&   0.16& $^{  +0.03}_{  -0.04}$ &    0.25& $\pm$   0.07&   0.11& $^{  +0.02}_{  -0.03}$ \\
 6312.10 & [S~{\sc iii}]   &    2.39& $\pm$   0.07&   1.47& $\pm$   0.04 &    0.92& $\pm$   0.10&   0.42& $\pm$   0.04 \\
 6461.95 & C~{\sc ii}      &    0.13& $\pm$   0.02&   0.08& $^{  +0.01}_{  -0.02}$  &    0.52& $\pm$   0.07&   0.22& $\pm$   0.03\\
 6548.10 & [N~{\sc ii}]    &   22.72& $\pm$   0.11&  13.22& $\pm$   0.07 &   34.66& $\pm$   0.30&  14.29& $\pm$   0.13 \\
 6562.77 & H~{\sc i}       &  478.84& $\pm$   0.66& 277.58& $\pm$   0.17 &  705.65& $\pm$   1.16& 289.19& $\pm$   0.35\\
 6583.50 & [N~{\sc ii}]    &   69.02& $\pm$   0.25&  39.81& $\pm$   0.15 &  106.91& $\pm$   0.53&  43.46& $\pm$   0.23 \\
 6678.16 & He~{\sc i}      &    7.34& $\pm$   0.07&   4.14& $\pm$   0.04 &   13.15& $\pm$   0.08&   5.15& $\pm$   0.03 \\
 6716.44 & [S~{\sc ii}]    &    9.49& $\pm$   0.08&   5.30& $\pm$   0.05  &   10.80& $\pm$   0.15&   4.17& $\pm$   0.06\\
 6730.82 & [S~{\sc ii}]    &   11.60& $\pm$   0.09&   6.46& $\pm$   0.05  &   13.74& $\pm$   0.16&   5.28& $\pm$   0.06\\
 7065.25 & He~{\sc i}      &    6.85& $\pm$   0.34&   3.54& $\pm$   0.18  &    3.18& $\pm$   0.38&   1.08& $\pm$   0.13\\
 7135.80 & [Ar~{\sc iii}]  &   26.68& $\pm$   1.39&  13.58& $\pm$   0.71  &   33.00& $\pm$   0.16&  10.94& $\pm$   0.06\\
 \hline
\end{longtable}
}

\end{document}